\documentclass[]{aastex631}

\usepackage{longtable}
\usepackage{upgreek}
\begin{document}

%\title{What Drives the Variability in Luminous Blue Variable Stars?}
\title{An Investigation into the Variability of Luminous Blue Variable Stars with \textit{TESS}%:\\ 
%Are there Differences in short-period $\alpha$ Cygni variations between Strong-Active, Weak-Active, and Dormant/Candidate LBVs?
}

\author[0000-0002-0786-7307]{Becca Spejcher}
\affiliation{Department on Physics and Astronomy, Embry-Riddle Aeronautical University, 
3700 Willow Creek Rd, 
Prescott, AZ 86301, USA}

\author[0000-0002-2806-9339]{Noel D. Richardson}
\affiliation{Department on Physics and Astronomy, Embry-Riddle Aeronautical University, 
3700 Willow Creek Rd, 
Prescott, AZ 86301, USA}

\author{Herbert Pablo}
\affiliation{American Association of Variable Star Observers, 185 Alewife Brook Pkwy., Cambridge, MA, USA}

\author{Marina Beltran}
\affiliation{Department on Physics and Astronomy, Embry-Riddle Aeronautical University, 
3700 Willow Creek Rd, 
Prescott, AZ 86301, USA}
\affiliation{Physics and Astrophysics Department, DePaul University, 2219 N. Kenmore Ave., Chicago, IL 60614, USA}

\author{Payton Butler}
\affiliation{Department on Physics and Astronomy, Embry-Riddle Aeronautical University, 
3700 Willow Creek Rd, 
Prescott, AZ 86301, USA}
\affiliation{BASIS Prescott, 1901 Prescott Lakes Parkway, Prescott, Arizona 86301, USA}

\author{Eddie Avila} %Edward?
\affiliation{Department on Physics and Astronomy, Embry-Riddle Aeronautical University, 
3700 Willow Creek Rd, 
Prescott, AZ 86301, USA}
\affiliation{BASIS Prescott, 1901 Prescott Lakes Parkway, Prescott, Arizona 86301, USA}

% \collaboration{20}{(AAS Journals Data Editors)}
% \affiliation{AAS Journals Associate Editor-in-Chief}

% \author{Amy Hendrickson}
% \altaffiliation{AASTeX v6+ programmer}
% \affiliation{TeXnology Inc.}

% \author{Julie Steffen}
% \affiliation{AAS Director of Publishing}
% \affiliation{American Astronomical Society \\
% 1667 K Street NW, Suite 800 \\
% Washington, DC 20006, USA}



\begin{abstract}
Luminous Blue Variables (LBVs) are enigmatic, evolved, massive stars. Their variability has been observed to be episodic with large eruptions, along with variations on time-scales of days to decades. We have extracted light curves of 37 LBVs from the first four years of the \textit{TESS} mission. These light curves provide two years of photometric time-series for stars in the LMC, with several months of data for Galactic or SMC targets. We analyze the Fourier properties of the stellar light curves to determine their characteristic frequencies and red noise amplitudes, comparing them to mass-loss parameters through H$\alpha$ strength, and in the case of the LMC stars, $B-V$ color and luminosity as estimated by their apparent $g$-magnitudes. We confirm the absence of correlation between any of the Fourier parameters and stellar parameters, implying that there is no trend in how these stars vary as measured with these photometric data, which may point towards these stars being an extension to the supergiant $\alpha$ Cygni variables and not a unique class of object with regards to their short-term variations. 
%This may point to LBVs being a standard stellar type in the upper H-R diagram without the need to classify them by their variability properties which appear to all be similar for the short-period $\alpha$ Cygni-type variations. 

\end{abstract}



\section{Introduction} \label{sec:intro}

Massive stars, while rare, provide one of the best means to study extreme physics in our Galaxy and Universe. They provide the vast amount of ionizing radiation in the Universe, send feedback into the interstellar medium via stellar winds and terminal supernova explosions, and are the seeds to the black holes observed throughout cosmic time through both electromagnetic and gravitational wave observations. It has been shown recently that a vast majority of massive stars are born in binary or higher-order systems, meaning that massive stars can undergo both traditional evolutionary paths or have strong binary interactions including merger events or Roche lobe overflow and spin-up \citep{2012Sci...337..444S}. 

About a half century ago, it was realized that the upper H-R diagram showed that stars that have luminosities in excess of $10^6 L_\odot$ are not observed to become red supergiants \citep{1979ApJ...232..409H}. The prevailing thought at the time was that these stars had instabilities that would prevent them from evolving to the cool effective temperatures and large radii of the red supergiants. This led to the standard view of the evolution of massive stars to include a post-main sequence stage of massive stars that would allow these stars to transition from a hydrogen-rich O star to a hydrogen-deficient Wolf-Rayet stars. This stage of evolution was deemed the luminous blue variable (LBV) stage which can include episodic eruptions such as those observed with P Cygni or $\eta$ Carinae along with high mass-loss rates \citep{1994PASP..106.1025H, 2001A&A...366..508V}.  

LBVs were well defined in their observational characteristics by \citet{2001A&A...366..508V} although the first review of these stars \citep{1994PASP..106.1025H} shows many of the same classification details. The typical absolute magnitude for a normal LBV is $M_V < -9.5$, but there may be some related or similar objects with an $M_V \approx -8$, which corresponds to luminosities in the range of $10^{5.5-6.5} L_{\odot}$. As a massive star runs out of hydrogen in its core and evolves beyond the main sequence, its atmosphere expands and cools towards the red supergiant stage. However, the more-luminous LBVs will encounter the atmospheric Eddington luminosity limit \citep[empirically known as the Humphreys-Davidson limit, see][]{1979ApJ...232..409H, 1984Sci...223..243H, 1988ApJ...324..279L} and are thus unable to become red supergiants, while the lower luminosity LBVs are thought to have been red supergiants \citep[e.g., HD160529;][]{1991A&A...247..383S}.

LBVs are subject to at least four types of photometric variability, which help fulfill the criterion of the ``variable" part of the name. The first, the ``giant eruptions" have amplitudes of $\ge 2$ magnitudes in the visual. This variability is extremely rare, and has only been observed in a few stars and is not a necessary criterion for classification as an LBV. In the Galaxy, only P Cygni and $\eta$ Carinae have been observed to exhibit such changes. The star HD 5980 in the SMC erupted in the 1990s \citep{koe10}. Other examples are often called ``supernova imposters" and are extragalactic in origin, and they are only seen in supernova surveys \citep{2011MNRAS.415..773S}. This large-scale variability is not well understood, both from a theoretical standpoint and an observational perspective because of the small number of observations of such events. 

The second type of variability is observed on timescales of approximately 10--40 years, where LBVs exhibit large ``eruptions", also called long S Doradus-phases \citep[long SD-phase][]{2001A&A...366..508V}. The term SD-phase originated with \citet{1997A&AS..124..517V} because the mass loss rate is not seen to vary dramatically during these changes. These variations are typically on the order of 0.5--2 magnitudes, and the star appears redder when brighter in the optical. One of the discoveries related to these variations is that the bolometric luminosity is roughly constant throughout this variability.

Similar to this, the third type of variability is the short-SD phase which \citet{1994PASP..106.1025H} refer to as ``oscillations", in which the stars display similar flux and color variations, but on timescales of less than 10 years (but usually at least 1 year in duration). It is unknown if the long-SD and short SD-phases are of different underlying origin, but an LBV can exhibit both simultaneously. The short-SD phase of P Cygni has an amplitude of $\triangle V \sim 0.05$ mag \citep{me-pcyg}.

Lastly, these stars are also subject to microvariations that occur on timescales of a few days to about a month. These are similar to instabilities ($\alpha$-Cygni variability) observed in other hot supergiants \citep[e.g. Deneb;][]{2011AJ....141...17R}. The amplitude is small, and these microvariations can be ignored when considering the long time scales and amplitudes of the long and short SD-phases or the great eruptions. These variations could be nonadiabatic strange mode radial pulsations that are driven by the $\kappa$ mechanism caused by iron-opacity variations. However, \citet{1998A&A...335..605L} argued that the timescales of the microvariations observed in LBVs were too long to be caused by strange modes. 

The variability types have led to LBVs being further classified based on their variability. For example, an LBV that shows drastic variability over long time-scales may be considered strong-active, while the LBVs with smaller amplitude variability are classified as weak-active. P Cygni, despite its historical eruptions, is considered weak-active as its variability is of very low-amplitude. Stars that show only $\alpha$ Cygni variability but reside in the same parts of the H-R diagram below the Humphreys Davidson limit but still with a high luminosity, or stars that have a spectrum similar to other LBVs, showing P Cygni-type profiles and/or strong H$\alpha$ emission are considered dormant or candidate LBVs.
%Luminous Blue Variables, or “LBVs”, are unstable, massive stars in the transition from being post-main sequence stars to being terminal stars with strong winds, such as Wolf-Rayet stars (Humphreys and Davidson 1994). 
%LBVs have events where the star loses several solar masses of material within the span of a few years. These are extreme eruptions and can be seen over large distances, because the energy output is similar to that of a supernova. It is unclear where in stellar evolution LBVs fit, but it is thought to be an important evolutionary phase of massive stars where they transition into hydrogen-free Wolf-Rayet stars. Understanding this phase of stellar evolution can help determine the star's fate and the type of supernova explosion that results from massive stars \citep{2019MNRAS.489.4378S}. 

% \subsection{Evolutionary State Theories}
% There are many theories as to what evolutionary phase LBVs fit into. One of the earliest theories was about how LBVs are a transitional phase for single stars shifting from hydrogen-rich to hydrogen-poor stars by their high mass-loss rate eruptions. Where \cite{2015MNRAS.447..598S} find LBVs to be a product of binary evolution. The main two theories are that LBVs are a result of a companion that exploded as a supernova or result of a stellar merger.

Classically, LBVs have been considered the transition between the main-sequence and the hydrogen-deficient Wolf-Rayet stars \citep{1994PASP..106.1025H}. This was challenged by some recent observations where LBVs or LBV-like stars have been seen to seemingly explode as type IIn supernovae \citep[e.g.,][]{2013MNRAS.430.1801M,2016MNRAS.463.3894E}. This has led 
% %\cite{1994PASP..106.1025H} suggested that LBVs are evolved massive stars because of their instability characterized by high luminosities, high mass-loss rates, and ejected elements like helium and nitrogen. They concluded with the idea that LBVs start as very massive stars with closer to average mass loss that evolves as massive stars normally do, moving to the right of the HR-diagram. They then become unstable at a temperature or radius due to some mechanism causing the stochastic behavior that leads to their characteristic high-mass loss rates. The stars veer toward the left of the HR-diagram because of the mass-loss, resulting in a temporary stability, and increasing the luminosity-mass ratio leads to an decrease of stability. Instability was said to recur as these stars evolve toward cooler temperatures until the mass-loss has formed a Wolf-Rayet (WR) star. 
% This theory has since not been widely accepted. Observations of the Type IIn supernova played a big role in unlinking the eruption behavior of LBVs to a transitional phase or high mass stars. Type IIn supernovae are white dwarfs of 8-10 solar masses (up to 40-50 solar masses) that erupt up to a solar mass years before a core-collapse that resemble eruptions of LBVs \citep{2018AJ....156..294A}.
\citet{2015MNRAS.447..598S} and \citet{2018AJ....156..294A} to approach the theory of LBV evolution from the angle of the isolation of LBVs. \citet{2015MNRAS.447..598S} analyzed the statistical component of the isolation of LBVs and have argued LBVs are a product of binary evolution. They found that because LBVs of the Galaxy and LMC are statistically more isolated than O-type stars and more so WR stars, LBVs are not likely to be a part of the transitional phase of a WR star. Instead, LBVs are mass gainers that are enriched and spun up by their companion and can be kicked out of their cluster because of this. {This was challenged by \citet{2016ApJ...825...64H} who examined the populations in the Local Group galaxies M31 and M33  as well as the LMC and SMC. \citet{2016ApJ...825...64H} found that the populations are not different than the O star population. }\citet{2018AJ....156..294A} argues that the sample of \citet{2015MNRAS.447..598S} was composed using the SIMBAD database to identify the closest known O-type star. \citet{2018AJ....156..294A} included the projected angular separation from the nearest bright blue stars neighbor for each of the three LMC samples of LBVs considered \citep[with samples defined by][]{2015MNRAS.447..598S, 2016ApJ...825...64H, 2018RNAAS...2..121R}. From this analysis, \citet{2018AJ....156..294A} found that LBVs are not as isolated as \citet{2015MNRAS.447..598S} found them to be and LBVs should be considered to be the transitional phase between the main-sequence and the terminal stages before SN explosions, although some stars seem to explode as supernovae while still appearing as LBVs. In support of \citet{2018AJ....156..294A}, \citet{2022A&A...657A...4M} find that LBVs in the Galaxy have a similar multiplicity as other OB stars and WR stars.

% \subsection{Brief History of P Cygni}
% P Cygni has been recorded twice to have massive eruptions in 1600 and 1654. 
% Past studies given from Elliott et al: ~What’s Important to note?~ \\
% Photometric studies began in earnest with Percy & Welch (1983), Percy et al. (1988), and de Groot (1990). These studies displayed three major timescales: a short period around 17 days associated with typical 훼 Cygni varibility, a ∼100 day period similar to other known LBVs, and a cycle of years that can be classified as a short SD-phase. These timescales were confirmed by de Groot et al. (2001a). Spectroscopically, the star has been studied most notably by Markova et al.(2001b) and Richardson et al. (2011). Markova et al. (2001b) comparedcpreviously noted 푈퐵푉 photometric data and new H훼 spectroscopy. These photometric data confirm the presence of a slow variation in brightness on a timescale of 7.4 years. The H훼 equivalent width determinations indicate the presence of a slow component, dubbed the Very-Long Term Component, in the variability of H훼, and is also a part of the variable SD-phase of the star. In recent years, the star has become a popular object for amateur astronomers to collect simultaneous spectroscopy and photometry, which has led to a possible detection of a 318 d period in the H훼 profile (Pollmann & Bauer 2012; Pollmann & Vollmann 2013; Pollmann 2016, 2020). From Elliott et al.

In recent years, asteroseismology has entered a Renaissance with space-based photometry missions such as \textit{MOST}, \textit{CoRoT}, \textit{Kepler} (and \textit{K2}), \textit{BRITE}-Constellation, and most recently \textit{TESS} \citep{2021RvMP...93a5001A}. 
Massive stars were investigated in depth by \citet{2011A&A...533A...4B}, \citet{2019A&A...621A.135B, 2019NatAs...3..760B, 2020A&A...640A..36B} and \citet{2020A&A...639A..81B} using data from several of these missions. The results of these findings showed that massive stars have a strong, low-frequency component in the Fourier domain, which has been interpreted to be a stochastic component of variability, {with an early reporting with space-based photometry from \citet{2011A&A...533A...4B} although some ground-based discoveries were seen before then}. The shape of this stochastic component indicated that the driving mechanism for the variability of the main-sequence OB stars is likely internal gravity waves, but in more evolved massive stars the physical mechanism(s) responsible remain unclear \citep[see ][for a review]{2023Ap&SS.368..107B}.

In the context of LBVs, very few have had an asteroseismic analysis performed. Recently, \citet{2022MNRAS.509.4246E} used a long-time series of precision photometry data from \textit{BRITE}-Constellation nanosatellites spanning from 2014-2019 to study the driving mechanism(s) of P Cygni, an LBV that is visually very bright and well studied \citep[see ][for a review on its variability]{me-pcyg}. \citet{me-pcyg} showed how P Cygni's H$\alpha$ profile's variations correlate with the short-SD phase, implying small shell-like ejections on a time-scale of a few years. \citet{2013ApJ...769..118R} were later able to spatially resolve the wind in the near-infrared to compare to the wind models from the non-LTE radiative transfer code CMFGEN. \citet{2022MNRAS.509.4246E} found that the star had no periodic behavior, and the observed stochastic variations were similar to those of the massive O stars, suggesting the driving mechanisms for P Cygni's variability with an arguably small red noise amplitude were possibly related to internal gravity waves in main sequence OB stars \citep{2020A&A...640A..36B} or due to sub-surface convection as investigated by \citet{2018Natur.561..498J}. One of the few other LBVs with precision photometric time-series analyzed in the literature was the enigmatic massive binary $\eta$ Carinae \citep{2018MNRAS.475.5417R}, where two years of data from \textit{BRITE}-Constellation show that $\eta$ Car may have tidally excited oscillations. {Recently, \citet{2021MNRAS.502.5038N} presented an analysis of a few LBVs with \textit{TESS}, finding that red noise was present in all eight stars examined. Their analysis included eight stars that are very isolated compared to other LBVs with little contamination from background stars.} 

With the large amount of publicly-available \textit{TESS} photometry, we used data from the first four years of the mission to investigate the morphology of the Fourier power spectra of a larger sample of LBVs and candidate LBVs. In this paper, we extend the previous work of \citet{2021MNRAS.502.5038N}, including more Sectors and especially considering non-Galactic LBVs in addition to Galactic ones. Such a combined sample of LBVs has not been examined with high-precision photometry previously allowing us to compare the variability properties of LBVs with characteristics of strong-, weak-, or dormant/candidate LBVs \citep[as discussed in][]{2001A&A...366..508V}. We discuss the \textit{TESS} observations and associated spectroscopy in Section 2 and then describe the Fourier calculations and morphology characterizations in Section 3. We compare these measurements and discuss the findings in Section 4 and then conclude this study with an outlook for future studies in Section 5.

\section{Observations and reductions}

%\subsection{Transiting Exoplanet Survey Satellite (TESS)}
%TESS [what is other important background info on TESS]
\textit{TESS} is a space telescope that was launched in 2018 to survey the sky with time-series precision photometry with the intention of detecting transiting exoplanets around nearby stars \citep{2015JATIS...1a4003R}. In order to find transiting planets, \textit{TESS} began its mission by monitoring the sky by centering its four cameras around the ecliptic pole to equator. Each sector is observed continuously for $\sim$1 month. With the LMC positioned near the southern ecliptic pole, this satellite galaxy was monitored continuously for the first and third years of the \textit{TESS} mission. The full-frame images from which we extracted the light curves were taken with a 30-minute cadence for the first two years of the \textit{TESS} mission, and with a 10-minute cadence in years 3 and 4. We limited our analysis to these years of data to create a dataset we could analyze without continually adding additional data. 

While \textit{TESS} is not ideal for monitoring stars in the LMC with its large 21\arcsec\ pixels, LBVs are some of the visually most luminous stars in our Galaxy and satellite galaxies. Furthermore, the stars have been considered somewhat isolated from nearby bright stars as discussed in \citet{2015MNRAS.447..598S}. While this is not necessarily true \citep{2018AJ....156..294A}, the LBVs will dominate the light curves that are extracted at the position of these stars due to their extreme brightness. The space craft lies in a highly elliptical orbit around Earth to provide a stable environment for precision photometry. The resulting orbital geometry means that every $\sim$13 d, the space craft transmits data back causing gaps in the time series. We present the sample used along with the \textit{TESS} observation properties in Table \ref{sample}, and discuss the observational history of the targets in our appendix.

% \begin{longrotatetable}
\begin{deluxetable}{lclcccccc}
\tablecaption{LBVs studied in this survey \label{sample}}
\tablehead{
\colhead{} 		& \colhead{HD} & \colhead{TESS} 	             & \colhead{Number of} & \colhead{$B-V$} 	& \colhead{$g$} & \colhead{Contamination}	& \colhead{LBV} 	& \colhead{}   \\
\colhead{Name} 	& \colhead{Number} & \colhead{Sectors} 	&   \colhead{sectors} 	& \colhead{(mag)} 	& \colhead{(avg. mag)}   & \colhead{(\%)}  & \colhead{type}   & \colhead{Population}	 \\
}
\startdata
HD 80077	&	80077	&	8-9, 35-36	&	4	&	1.29	&	9.14	&  2.93  &	c	&	G	\\
HR Carinae	&	90177	&	36-37	&	2	&	1.66	&	9.34 &  1.44	&	s-a	&	G	\\
AG Carinae & 94910  & 10-11	&	2 & 0.61	& 8.73 &  1.97  & s-a & G  \\
WRAY 19-46	&	148937	&	12, 39	&	2	&	0.41	&	12.53 &  10.38	&	c	&	G	\\
$\zeta^1$ Sco	&	152236	&	12, 39	&	2	&	0.52	&	8.65	& 0.36  &	d	&	G	\\
HD 160529	&	160529	&	39	&	1	&	1.21	&	9.61	&  0.00 &	s-a	&	G	\\
P Cygni & 193237 & 14-15, 41  & 3 & 0.42	& 8.16 &   0.03   & w-a & G \\
% Hen 3-519	&		&	10-11	&	2	&	0.79	&	11.37  &  30.39 	&	d	&	G	\\
Sher 25	&		&	10-11, 37	&	3	&	1.36	&	10.93	&  0.90  &	c	&	G	\\
WRAY 15-751	&		&	10-11, 37	&	3	&	2.13	&	12.53  &   0.30	&	s-a	&	G	\\ \hline
S Dor	&	35343	&	1-13 27, 29-37, 39	&	24	&	0.14	&	9.72  &   16.51 	&	s-a	&	L	\\
HD 34664	&	34664	&	1, 3-11, 13, 27-28, 30-31, 33-39	&	22	&	0.106	&	11.71  &  0.42	&	w-a	&	L	\\
HD 37836	&	37836	&	1-3, 5-6, 8-13, 27-33, 35-39	&	23	&	0.044	&	10.47	&   0.23   &	w-a	&	L	\\
HD 37974	&	37974	&	1-3, 5-6, 8-3, 27-33, 35-39	&	23	&	0.13	&	10.93	&  0.44   &	c	&	L	\\
HD 38489	&	38489	&	1-6, 8-13, 28-36, 38-39	&	23	&	0.365	&	11.89	&  2.42  &	w-a	&	L	\\
R 66	&	268835	&	1-7, 9-13, 27-37, 39	&	24	&	0.14	&	10.63	&   0.21   &	c	&	L	\\
R 71	&	269006	&	1-13, 27-30, 32-39	&	25	&	0.05	&	9.3	&   0.03   &	s-a	&	L	\\
R 78	&	269050	&	1-4, 6-13, 27-34, 36-39	&	24	&	0.041	&	11.56	&   0.16   &	c	&	L	\\
HD 269216	&	269216	&	1-13, 27, 29-37, 39	&	24	&	-0.283	&	10.03	&   0.19   & 	s-a	&	L\\
% R 84	&	269227	&	1-13, 27, 29-37, 39	&	24	&	0.062	&	11.58	&  24.67   &	c	&	L		\\
R 85	&	269321	&	1-13, 27, 29-37, 39	&	24	&	0.09	&	10.53	&  18.82  &	w-a	&	L	\\
R 99	&	269445	&	1-4, 6-13, 27-34, 36-39	&	24	&	0.52	&	11.58 &  0.35 	&	w-a	&	L	\\
HD 269582	&	269582	&	2-10, 12-13, 27, 29-39	&	23	&	-0.275	&	10.84 &    0.63 	&	s-a	&	L \\
HD 269604	&	269604	&	2-10, 12-13, 27, 29-37, 39	&	22	&	0.13	&	10.8 &   0.27	&	c	&	L	\\
R 110	&	269662	&	2-10, 12-13, 27-30, 32-39	&	23	&	0.24	&	10.62 &    0.16 	&	s-a	&	L	\\
SK -69 175 & 269687 & 2-13, 27-39 & 23 & $-$0.09 & 11.74 &  0.87  & d & L  \\
HD 269700	&	269700	&	2-7, 9-10, 12-13, 27, 29-37, 39	&	21	&	0.02	&	10.48 &   0.21	&	s-a	&	L	\\
% R 127	&	269858	&	1-3, 5-6, 8-13, 27-33, 35-39	&	23	&	0.09	&	9.57 &   83.49	&	s-a	&	L\\
HD 269859	&	269859	&	1-3, 5-6, 8-13, 27-33, 35-39	&	23	&	0.031	&	10.2 &   16.56	&	c	&	L	\\
R 74	&	268939	&	1-13, 27-30, 32-39	&	25	&	0.04	&	10.99 &   0.11 	&	w-a	&	L	\\
CPD-69 500	&		&	1-6, 8-13, 28-36, 38-39	&	23	&	0	&	11.89 &    0.75  	&	c	&	L	\\
LHA 120-S 18	&		&	1-4, 6-9, 10-13, 27-34, 36-39	&	24	&	0.157	&	11.8 &   1.81	&	c	&	L	\\
LHA 120-S 61	&		&	1, 3-4, 6-7, 9-13, 27, 29-31, 33-34, 36-37, 39	&	19	&	-0.09	&	11.88 &    0.06 	&	d	&	L	\\
R 143	&		&	1-3, 5-6, 8-13, 27-33, 35-39	&	23	&	0.436	&	11.41 &   6.80 	&	s-a	&	L\\
% R 149	&		&	1-6, 8-13, 28-36, 38-39	&	23	&	-0.075	&	11.41 &    82.55 	&	w-a	&	L	\\
SK -69 279	&		&	1-6, 8-13, 28-36, 38-39	&	23	&	-0.055	&	12.74 &  1.63 	&	d	&	L	\\ \hline
R 40	&	6884	&	1-2, 27-28	&	4	&	0.12	&	9.76 &     0.20 	&	s-a	&	S	\\
\enddata
\tablecomments{The abbreviations for the populations are G = Galactic, S = SMC, and L = LMC, which we have shown in the three sub-sections interrupted by horizontal lines. For the LBV type, w-a is weak-active, s-a is strong-active, d is dormant, and c is for candidate. For bright stars, ASAS-SN is unreliable for the $g$ magnitudes and is not listed.}
\end{deluxetable}  
% \end{longrotatetable}

\begin{figure}[ht!]
\includegraphics[width=0.5\columnwidth]{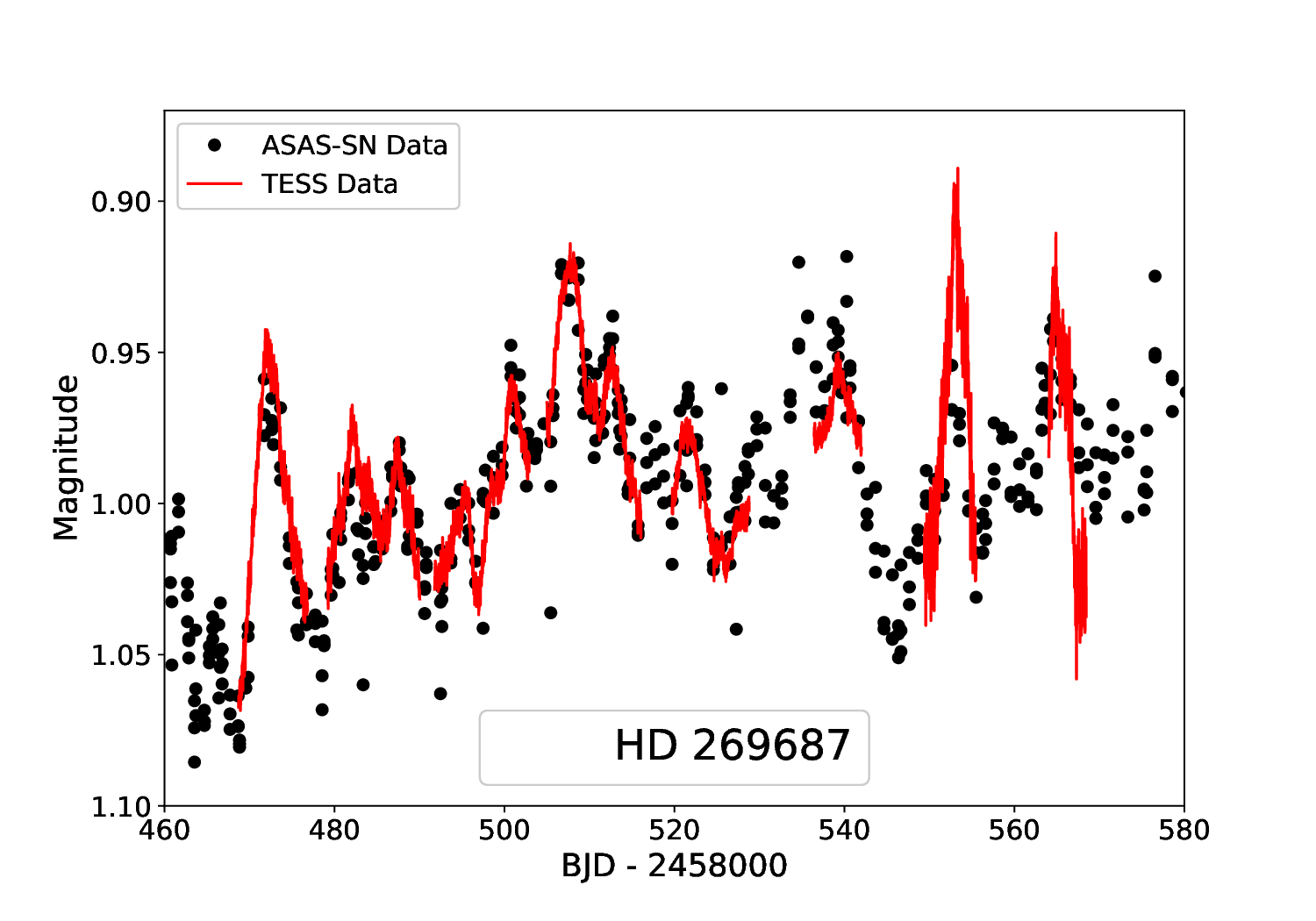}
\includegraphics[width=0.5\columnwidth]{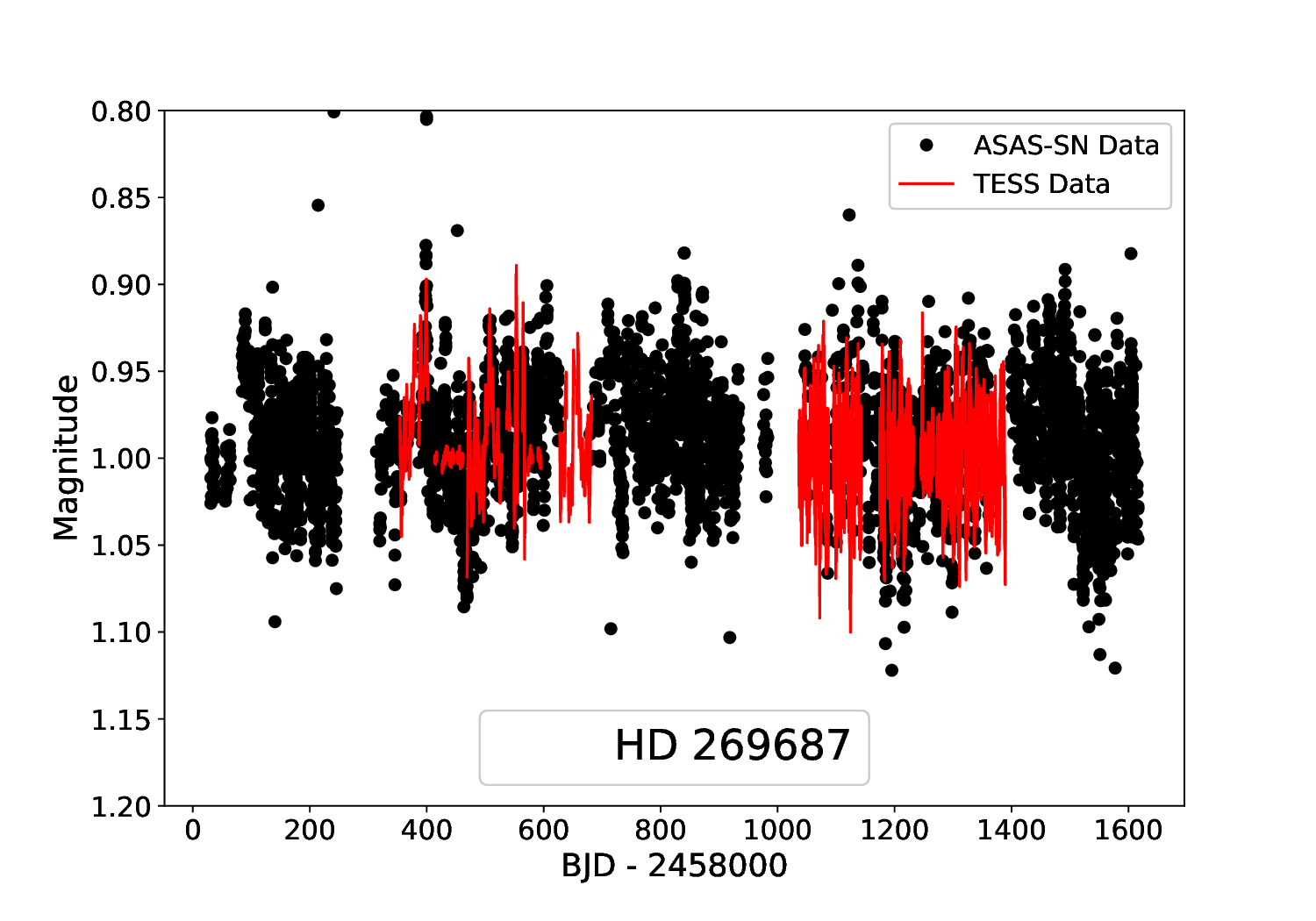}
\includegraphics[width=0.5\columnwidth]{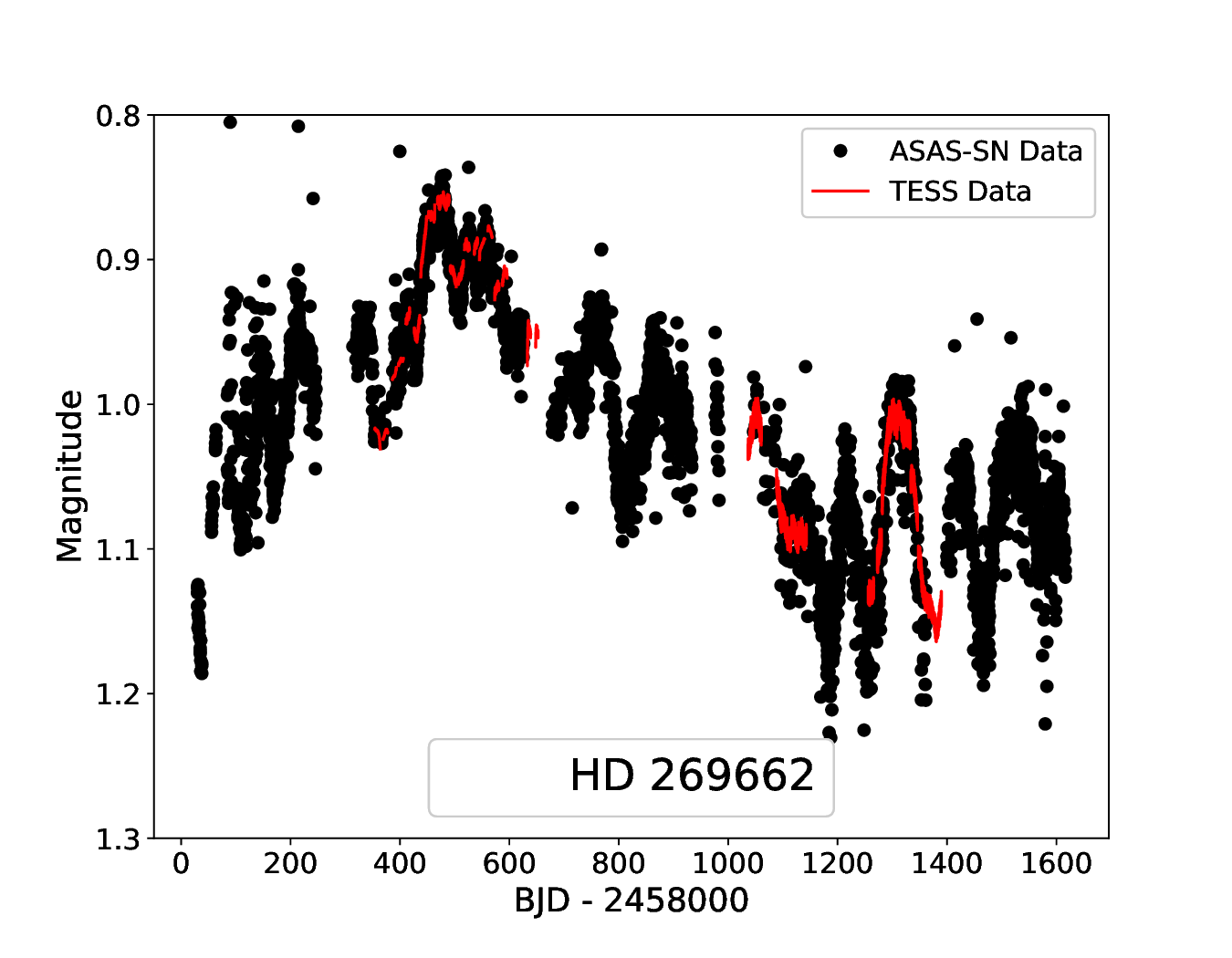}
\caption{%This graph shows the comparison of TESS sectors 6-9 and ASAS-SN for HD 269687.
We show the light curve of HD 269687 in the top two panels, both over a small three-sector time frame (top left) and over the full time frame of our analysis (top right). To show that our reduction of the light curves works well both for the weak-active or dormant/candidate LBVs as well as for the strong-active LBVs, we show the strong-active LBV HD 269662's light curve in the bottom frame. In all panels, the black points represent the ASAS-SN ground-based photometry while the red points are the \textit{TESS} measurements.
\label{ASASSN}}
\end{figure}

For each object listed in the recent census of the LBVs in the Local Group by \citet{2018RNAAS...2..121R} that was bright enough for \textit{TESS} to have observed, we used the {\tt eleanor} package \citep{2019PASP..131i4502F} to extract the light curves of the objects in question. {We also used the SIMBAD database to examine how much contamination from neighboring stars could impact the light curves, which is included in Table \ref{sample}. If the star was in a crowded region with more than a 20\% flux contamination, we removed the star from our sample.} The resulting light curves provided several types of extractions, including ones that were de-trended that removed the long-term physically relevant portions of the light curve. This implied that we needed to correctly choose the best light curve extraction to probe the underlying astrophysics. Occasionally, the mid-sector timing for data downloads introduced artifacts into the light curve. We utilized the light curves that retained the intrinsic {``raw"} counts from the stars while removing instrumental artifacts from the {\tt eleanor}-extracted light curves, meaning that each sector had to be matched to other sectors as the individual sectors did not necessarily have the same flux measured in sequential sectors. {These raw counts were still corrected for variations in the background flux but keep the long-term variations intact for our analysis.} To ensure that we had astrophysically correct light curves, we compared these to the ground-based ASAS-SN photometry \citep{2014ApJ...788...48S, 2023arXiv230403791H}. {This allows us to have a light curve that includes all astrophysical information included ranging from the short-period $\alpha$ Cygni variations to the long-term S Doradus variability.} In Fig.~\ref{ASASSN}, we show comparison light curves for two LBVs. The first panel shows a three-sector region of the light curve of HD 269687 (SK $-$69 175). Over this three-month region of time, our \textit{TESS}-extracted light curve matches the ground-based ASAS-SN $g$-band light fairly well. We show the full light curve of this star in the second panel of Fig.~\ref{ASASSN}. HD 269687 is a candidate or dormant LBV, so we also show the light curve of a strong-active LBV (HD 269662; R 110) in Fig.~\ref{ASASSN}, showing that the reduction techniques work for both weak-active and strong-active LBVs. We also note that the ASAS-SN data are taken with a $g$-band filter, while the \textit{TESS} filter is a wide filter between $\sim$6000\AA\ and 1$\upmu$m.

In addition to our photometric time-series, we obtained snap-shot spectra of many of our targets with the CTIO 1.5 m telescope and the CHIRON spectrograph \citep{2013PASP..125.1336T}. The spectrograph was operated in the ``fiber" mode, resulting in a resolving power of $\approx$27,000. We aimed to have a signal-to-noise ratio (SNR) of at least $\approx 50$ in the continuum to measure the H$\alpha$ strength of each target as a proxy for the mass-loss rate. We tabulate the observing log for these observations in Table \ref{params} along with equivalent widths of H$\alpha$.

\section{The Fourier Properties of the LBV Light Curves}

We calculated the Fourier transform of the \textit{TESS} light curves using the {\tt period04} software platform \citep{2005CoAst.146...53L}. {These light curves include all astrophysical variability from the short-term $\alpha$ Cygni variations to the long-term S Doradus variations.} This examination of the data in this manner led us to reject a few sources from the LBV catalog of \citet{2018RNAAS...2..121R}, namely HD5980, HD 269128 (R81) and HD 326823, all of which are confirmed, known binaries \citep{2014AJ....148...62K,2002A&A...389..931T,2011AJ....142..201R} which have periodic signals that dominate the light curves. {We note that {\tt period04} does not handle gaps in the time-series, which does provide some issues with our Fourier properties, but these Fourier properties should be intercomparable for the population we are examining. }

% \subsection{Period04}
% Period04 is a software designed for astronomical data analysis. It focuses primarily on statistical analysis of time series data that contains gaps, which is why we can use it on our data. The program has a Fourier transform module that is based off a discrete Fourier transform and extracts new frequencies from the data. We used this module to compute Fourier transforms on each of our corrected light curves. This allowed us to find the peak frequency and initial amplitude of our data.

Recent advances in massive star asteroseismology (see review by \citet{2020FrASS...7...70B}) have used the characterization of the Fourier transform amplitude as a function of frequency with the form of
% \subsection{Nonlinear Regression}
% After each Fourier transform, we downloaded the data and read it into Python to start analyzing the data. We graphed the data in loglog and then fit the Fourier spectrum to the equation used by \citet{2022MNRAS.509.4246E}:
\begin{equation}
    \alpha_{\nu} = \frac{\alpha_0}{1+(\frac{\nu}{\nu_{char}})^{\gamma}} + C_w.
\end{equation}
{This is based on the characterization of the Fourier properties of the Sun as described by \citet{1985ESASP.235..199H}, \citet{2014A&A...570A..41K} and others}. Here, the amplitude of the Fourier transform $\alpha_\nu$ is a function of frequency $\nu$. $\alpha_0$ represents the amplitude of the semi-Lorentzian fit at a frequency value of zero, $\nu_{char}$ represents a characteristic frequency, $\gamma$ represents the logarithmic amplitude gradient, and $C_w$ represents the white noise of the data. {While this formulation actually applies to the power of the variations, not the amplitude, recent asteroseismic studies based on space photometry have typically fit amplitudes, which we do here in that tradition.}

\begin{figure}[h!]
\includegraphics[width=0.95\columnwidth]{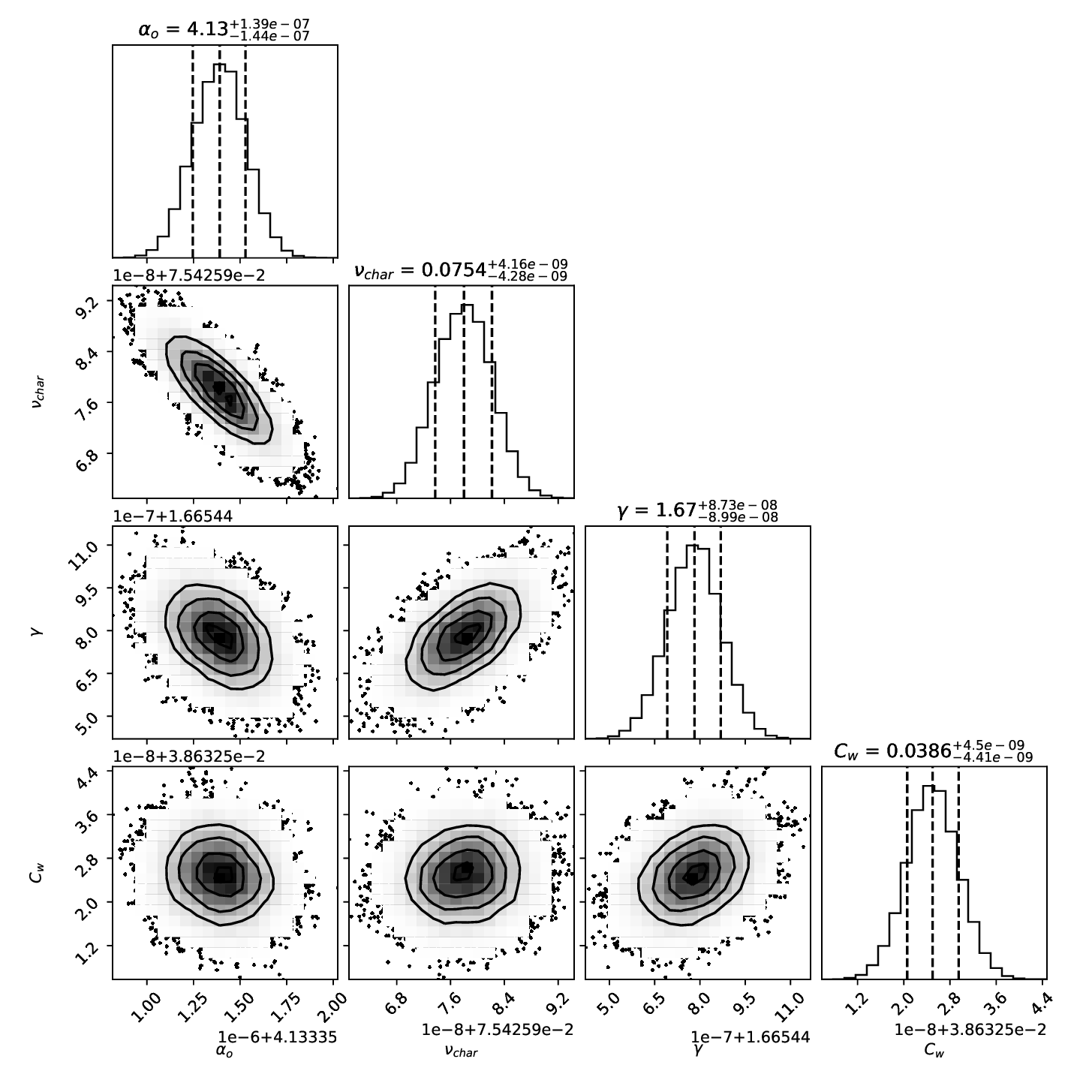}
\caption{Corner plot produced from the MCMC simulation of S Doradus. The graph gives the parameters and correlation between each parameter.
\label{fig: Corner Plot}}
\end{figure}

We began our characterization of the Fourier transformation with the python package {\tt lmfit}\footnote{https://lmfit.github.io/lmfit-py/} that uses nonlinear regression to fit the Fourier transform of each star with the given parameters. We visually matched these Fourier transforms to the model equation given above and then used those values as initial parameters for {\tt lmfit}. This fitting routine then gives parameters and a corresponding error that was calculated automatically by inverting the second derivative of the matrix of the parameter, which is similar to a goodness-of-fit method.

\begin{figure}[ht!]
\centering
\includegraphics[width=0.5\columnwidth]{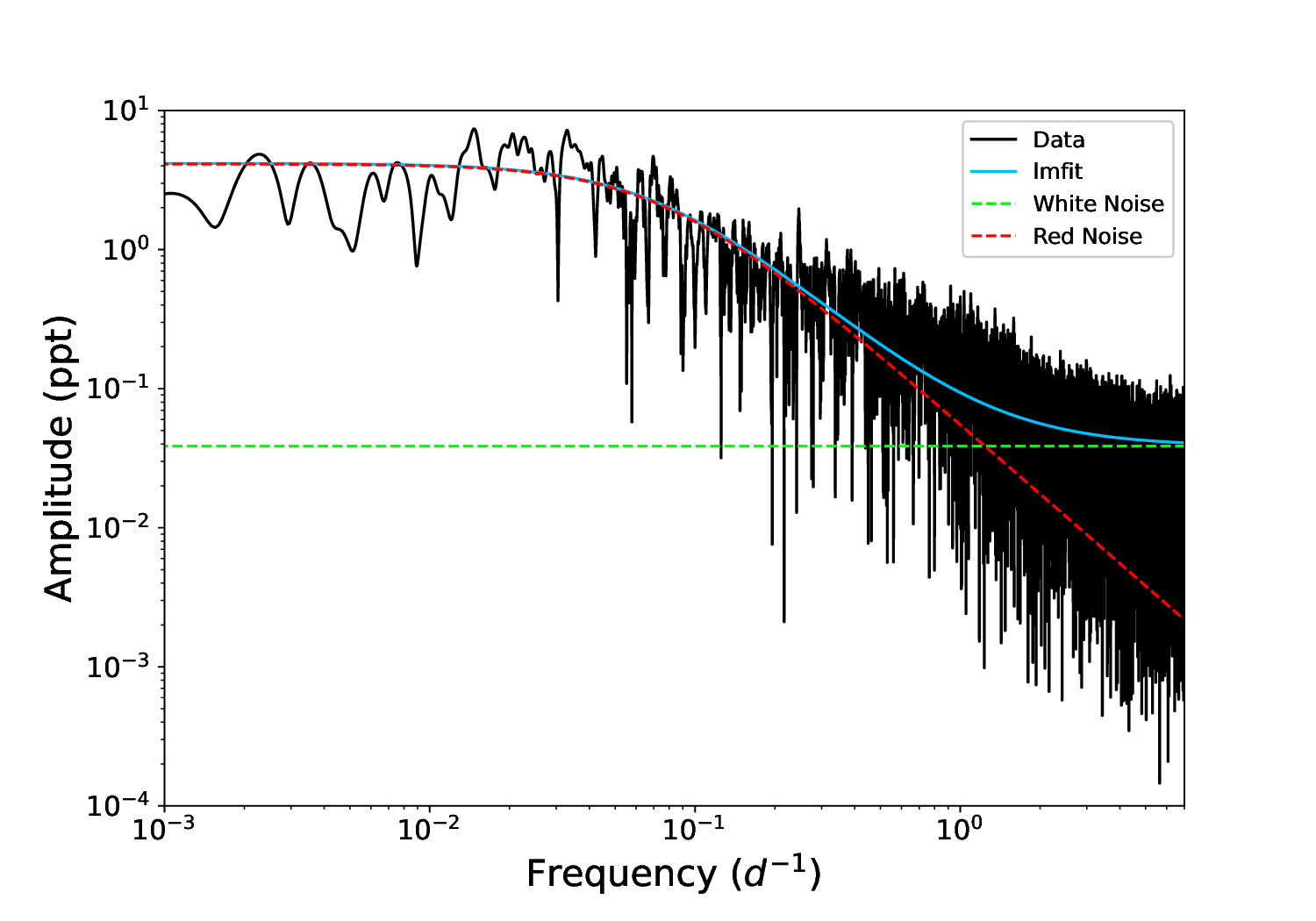}
\includegraphics[width=0.5\columnwidth]{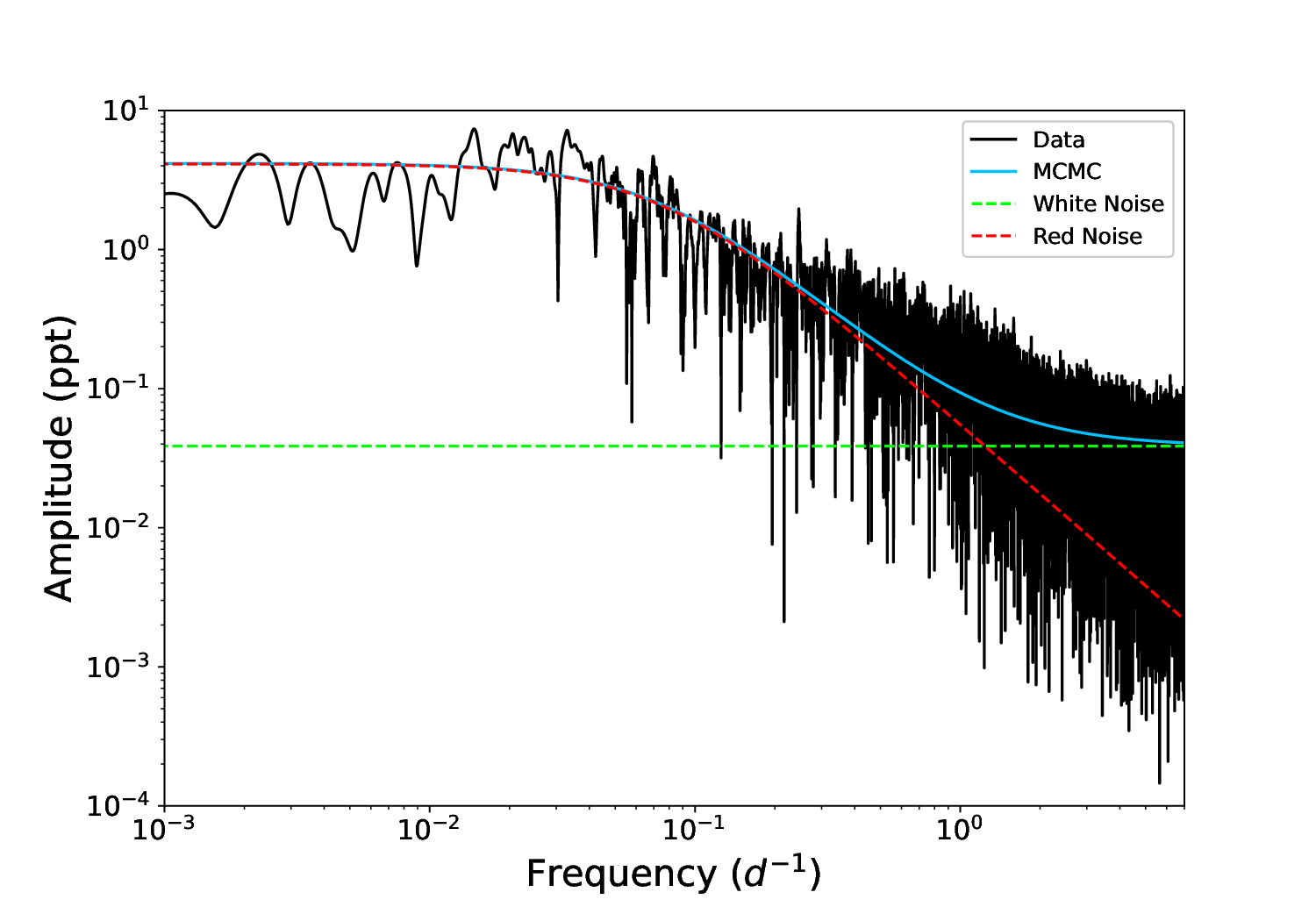}
\caption{Fourier transform, red noise, white noise ($C_W)$, and fit of $\alpha_\nu$. The top graph shows $\alpha_\nu$ using parameters given from lmfit and the bottom graph shows $\alpha_\nu$ using the parameters given from the Monte Carlo simulation. The parameters for each star can be found in Table \ref{params}.
\label{fig: S Dor fit}}
\end{figure}

To better sample the parameter space of the fits and search for correlations between parameters, we then used the {\tt emcee} package \citep{2019JOSS....4.1864F} that samples the parameter space based on the output of the {\tt lmfit} fitting routine. The {\tt emcee} package uses a Markov Chain Monte Carlo approach to sample the parameter space. We typically used 15 walkers and 10000 steps for each run to allow the sampler to converge on a best-fit value for each parameter. We could then visualize the fit with the python package {\tt corner}\footnote{https://corner.readthedocs.io/en/latest/}. To ensure that the fits were realistic, we used a Fourier transform of the BRITE-Constellation data from \citet{2022MNRAS.509.4246E} to compare our results to those presented by \citet{2022MNRAS.509.4246E}. We found our method produced similar results to \citet{2022MNRAS.509.4246E} and then examined each individual star in our sample from \textit{TESS}. An example of the fit of the Fourier transform of S Doradus is shown in Fig.~\ref{fig: Corner Plot}. The corresponding fit to the Fourier transform is shown in Fig.~\ref{fig: S Dor fit}. The parameters of all of the fits are given in Table \ref{params}. The MCMC sampler produced reliable results but also had very small errors, often several orders of magnitude smaller than the values. {To find more realistic errors, which are included in Table~\ref{params}, we used the same technique as \citet{2021MNRAS.502.5038N}. This technique uses the number of data points and the $\chi^2$ value for the fit where the actual error is estimated to be the error from the emcee fit multiplied by $\chi^2({\rm best fit})/(0.5\times N_{\rm data} - 4)$. These errors are given in Table~\ref{params}.}

%We anticipate the errors are more likely to be on the order 5--10\% so we did not include them in Table \ref{params}, but these values represent reasonable estimates of the parameters for the sample's Fourier properties. 

% Using emcee and corner, we were able to run Monte Carlo simulations for each star to find more accurate parameters and errors for the stars. We started by using the data P Cygni to make sure our notebook produced similar values as \citet{2022MNRAS.509.4246E}. Once the code was working, we ran the code for each set of data. The parameters and their errors for each star can be found in Table 1.

\begin{deluxetable}{lccccc}
\tablecaption{Fourier parameters of the LBVs studied in this survey \label{params}}
\tablehead{
\colhead{} 		& \colhead{Amplitude} & \colhead{Char. Frequency} 	             & \colhead{} & \colhead{White noise}  &
 \colhead{} \\
\colhead{Name} 	& \colhead{ ($a_0$) [ppt]} & \colhead{($\nu_{char}$) [d$^{-1}$]} 	&   \colhead{$\gamma$} 	& \colhead{($C_W$) [ppt]}   &   \colhead{$W_\lambda ({\rm H}\alpha )$ [\AA]}  \\ 
}
\startdata
P Cygni         & 10.5 $\pm$ 0.036 & 0.076 $\pm$ 0.0004 & 2.18 $\pm$ 0.015 & 0.0604 $\pm$ 0.001 \\
P Cygni (BRITE) & 5.99  & 0.078  & 1.88 & 0.250  &  \\
P Cygni (E22)   & 12.8 $\pm$ 0.191 & 0.033 $\pm$ 0.001  & 1.04 $\pm$ 0.009 & 0.288 $\pm$ 0.004 & \\
HD 269687	&	2.74 $\pm$ 0.013  &	0.220 $\pm$ 0.001	&	3.02 $\pm$ 0.033	&	0.103 $\pm$ 0.0006 & \\
AG Carinae *	&	3264.33 $\pm$ 98.9	& 0.0056 $\pm$ 0.0004 & 0.986 $\pm$ 0.0002	&	10.1 $\pm$ 0.630 & -78.52	\\
CPD-69 500	&	4.94 $\pm$ 0.057	&	0.034 $\pm$ 0.0009	&	1.05 $\pm$ 0.009	&	0.066 $\pm$ 0.0007	& -3.54 \\
HD 6884	&	152.24 $\pm$ 1.82	&	0.0072 $\pm$ 0.0002	&	1.16 $\pm$ 0.013	&	0.919 $\pm$ 0.01 & -2.86  \\
HD 37836	&	1310000 $\pm$ 1904.886	&	0.0455 $\pm$ 0.0001 	&	2.067 $\pm$ 0.0066	&	5603.19 $\pm$ 33.8	& -145.51 \\
HD 37974	&	6.19  $\pm$ 0.114	&	0.0089 $\pm$ 0.0004	&	0.944 $\pm$ 0.0093	&	0.028 $\pm$ 0.0006	 &  -447.46  \\
HD 38489	&	6.26 $\pm$ 0.038	&	0.0097 $\pm$ 0.0001	&	0.973 $\pm$ 0.0034	&	0.031 $\pm$ 0.0003	 & -547.67  \\
HD 80077 *	&	61.7 $\pm$ 0.056	&	0.0089 $\pm$ 0.00002 	&	1.006 $\pm$ 0.0005	&	0.223 $\pm$ 0.0003 & -6.50  \\
% HR Car *	&	2431.67	&	0.0063	&	1.1326	&	1.9149	\\
HD 160529 *	&	12.16 $\pm$ 0.544	&	0.061 $\pm$ 0.0033	&	3.85 $\pm$ 0.76	&	0.0508 $\pm$	0.015  & -7.98  \\
HD 269662	&	23.68 $\pm$ 0.158	&	0.143 $\pm$ 0.0015	&	1.90 $\pm$ 0.025	&	0.615  $\pm$ 0.025 & -15.08  \\
HD 269700	&	4.71 $\pm$ 0.020	&	0.086 $\pm$ 0.0007	&	1.52 $\pm$ 0.0007	&	0.078 $\pm$ 0.0003 &  -7.72  \\
% Hen 3-519 *	&	21.7 $\pm$ 0.892	&	0.045 $\pm$ 0.0034	&	1.39 $\pm$ 0.063	&	0.208 $\pm$ 0.022 & -227.79  \\
HR Carinae *	& 8.96 $\pm$ 0.426  &	0.231 $\pm$ 0.016 & 1.84 $\pm$ 0.148 &	0.089 $\pm$ 0.018	 & -54.51 \\
LHA 120-S 61	&	80.9 $\pm$ 8.151	&	0.297 $\pm$ 0.0058	&	1.22 $\pm$ 0.014	&	1.79 $\pm$ 0.044 & -64.47 \\
Sher 25 *	&	35.3 $\pm$ 1.192	&	0.013 $\pm$ 0.001	&	1.08 $\pm$ 0.0025	&	0.175  $\pm$ 0.001 & -3.05  \\
WRAY 15-751 *	&	250.3 $\pm$ 1.083	&	0.014 $\pm$ 0.0001	&	1.18 $\pm$ 0.0048	&	0.805 $\pm$ 0.0103 & -177.10 \\
WRAY 19-46 *	&	1339.99	$\pm$ 2.435 &	0.029 $\pm$ 0.0001	&	1.23 $\pm$ 0.002	&	5.43 $\pm$ 0.025 & -0.62  \\
$\zeta^1$ Sco *	&	14.2 $\pm$ 1.254 &	0.052 $\pm$ 0.001	&	1.21 $\pm$ 0.0086	&	0.157 $\pm$ 0.0013  & -8.76  \\
% HD 269858	& 77.9 $\pm$ 0.230	& 0.0066 $\pm$ 0.0004	&	1.19 $\pm$ 0.0036	&	0.109 $\pm$ 0.0016	 & -253.44  \\
HD 269859	&	79.9 $\pm$ 0.636	&	0.0013 $\pm$ 0.0002	&	0.841 $\pm$ 0.0029 	&	0.190 $\pm$ 0.0016	 & -4.27  \\
R 143	&	10.3 $\pm$ 0.037	&	0.029 $\pm$ 0.0002	&	1.34 $\pm$ 0.005	&	0.057 $\pm$ 0.0005	 & -5.57  \\
% R 149	&	10.4 $\pm$ 0.048	&	0.018 $\pm$ 0.0002	&	1.16 $\pm$ 0.0043	&	0.053 $\pm$ 0.0004 	& \\
SK -69 279	&	5.98 $\pm$ 0.059	&	0.044 $\pm$ 0.001	&	1.19 $\pm$ 0.01	&	0.083 $\pm$ 0.0007	&  \\
HD 268835	&	6.02 $\pm$ 0.124	&	0.033 $\pm$ 0.0014	&	1.29 $\pm$ 0.027	&	0.047 $\pm$ 0.0014 	 & -73.96  \\
HD 269006	&	100.8 $\pm$ 0.360	&	0.0028 $\pm$ 0.0002	&	1.09 $\pm$ 0.0039	&	0.169 $\pm$ 0.0014	&   \\
HD 268939	&	16.01 $\pm$ 0.064   &	0.023 $\pm$ 0.0019	&	1.21 $\pm$ 0.0046	&	0.058 $\pm$ 0.0008	 & -33.02  \\
LHA 120-S 18	&	14.7 $\pm$ 0.233	&	0.120 $\pm$ 0.0038	&	1.46 $\pm$ 0.025	&	0.167 $\pm$ 0.0046	&  \\
HD 269050	&	3.44 $\pm$ 0.021	&	0.014 $\pm$ 0.0016	&	1.52 $\pm$ 0.011	&	0.069 $\pm$ 0.0004	& \\
HD 269128	&	15.3 $\pm$ 0.049	&	0.089 $\pm$ 0.0005	&	1.85 $\pm$ 0.011	&	0.145 $\pm$ 0.0011	 & -29.93  \\
HD 269216	&	51.4 $\pm$ 0.310   &	0.012 $\pm$ 0.0002 	&	1.17 $\pm$ 0.0064	&	0.132 $\pm$ 0.0024	 &  -38.98  \\
HD 34664	&	63.06 $\pm$ 0.407	&	0.0015 $\pm$ 0.0015 	&	0.858 $\pm$ 0.0026 	&	0.138 $\pm$ 0.0011  & -817.24  \\
% HD 269227	&	7.94 $\pm$ 0.052	&	0.130 $\pm$ 0.0013	&	2.13 $\pm$ 0.038	&	0.181 $\pm$ 0.0025	& \\
HD 269321	&	18.9 $\pm$ 0.085	&	0.018 $\pm$ 0.0002	&	1.23 $\pm$ 0.0056	&	0.045 $\pm$ 0.0011	 & -12.12  \\
S Doradus	&	4.13 $\pm$ 0.020	&	0.075 $\pm$ 0.0006 	&	1.67 $\pm$ 0.014	&	0.039 $\pm$ 0.0007	 & -23.73  \\
HD 269445	&	4.28 $\pm$ 0.030	&	0.054 $\pm$ 0.0008	&	1.11 $\pm$ 0.0058	&	0.048 $\pm$ 0.0004	 & -30.53  \\
HD 269582	&	134.6 $\pm$ 0.790	&	0.0027 $\pm$ 0.0004 &	0.961 $\pm$ 0.0035	&	0.357 $\pm$ 0.0028 	&  \\
HD 269604	&	5.66 $\pm$ 0.063	&	0.041 $\pm$ 0.001	&	1.15 $\pm$ 0.011	&	0.059 $\pm$ 0.0014	& \\
\enddata
\tablecomments{For P Cygni, we include our analysis of the BRITE data reported by \citet{2022MNRAS.509.4246E} along with their measurements, indicated with (BRITE) and (E22) respectively.}
\end{deluxetable}

% \begin{figure}[ht!]
% %\centering
% \includegraphics[width=1.15\columnwidth]{ParameterGraphs.eps}
% \caption{This graph shows the amplitude (left), characteristic frequency (center), and logarithmic equivalent widths for each LBV. The type of LBV is shown by dot color --- red for strong-active, blue for weak-active, and black for candidate/dormant. The open dot shows the values that \citet{2022MNRAS.509.4246E} got for their values of P Cygni.
% \label{fig:amp-freq-ha}}
% \end{figure}

\begin{figure}[ht!]
%\centering
\includegraphics[width=0.8\columnwidth]{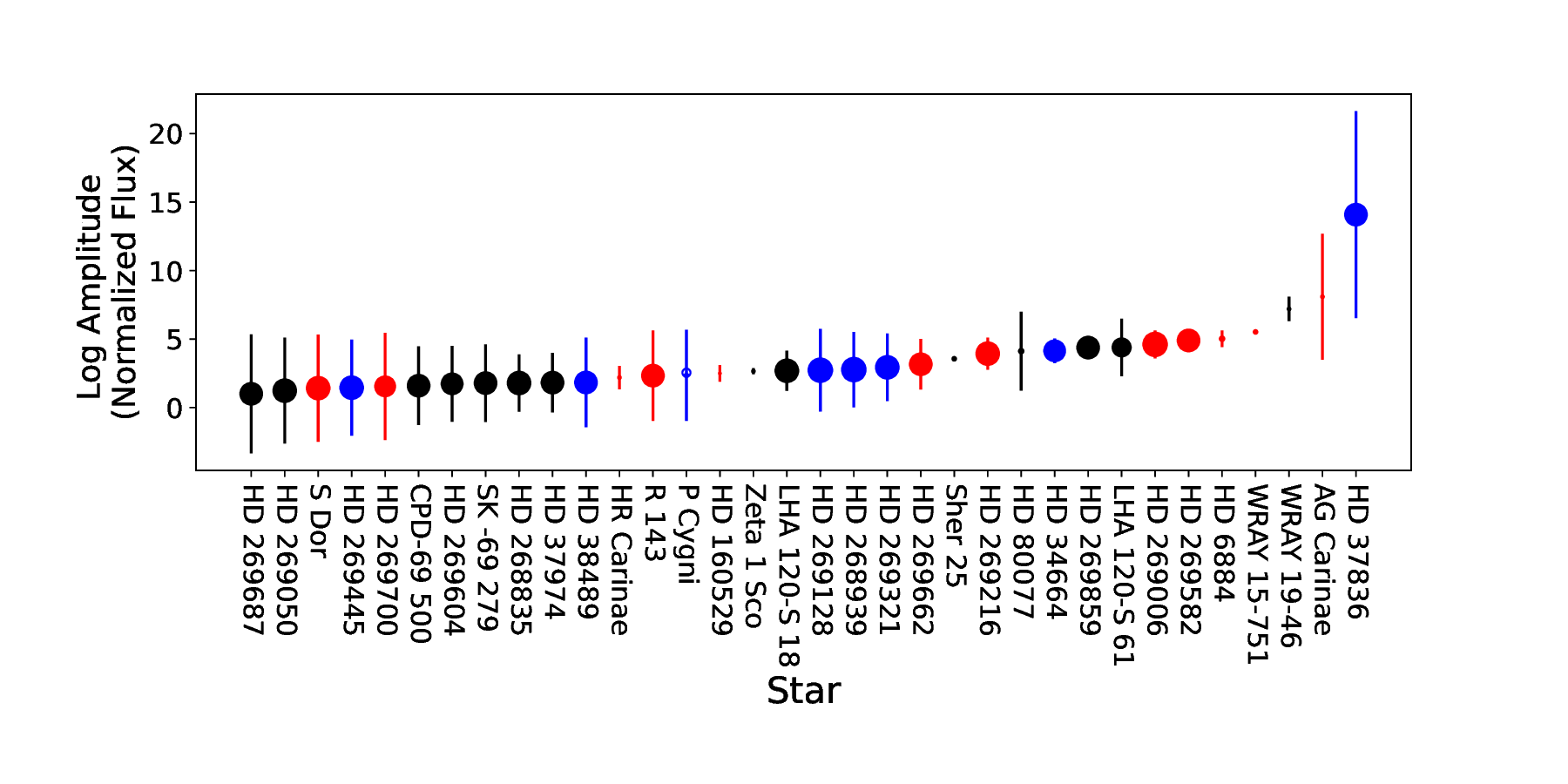}
\includegraphics[width=0.8\columnwidth]{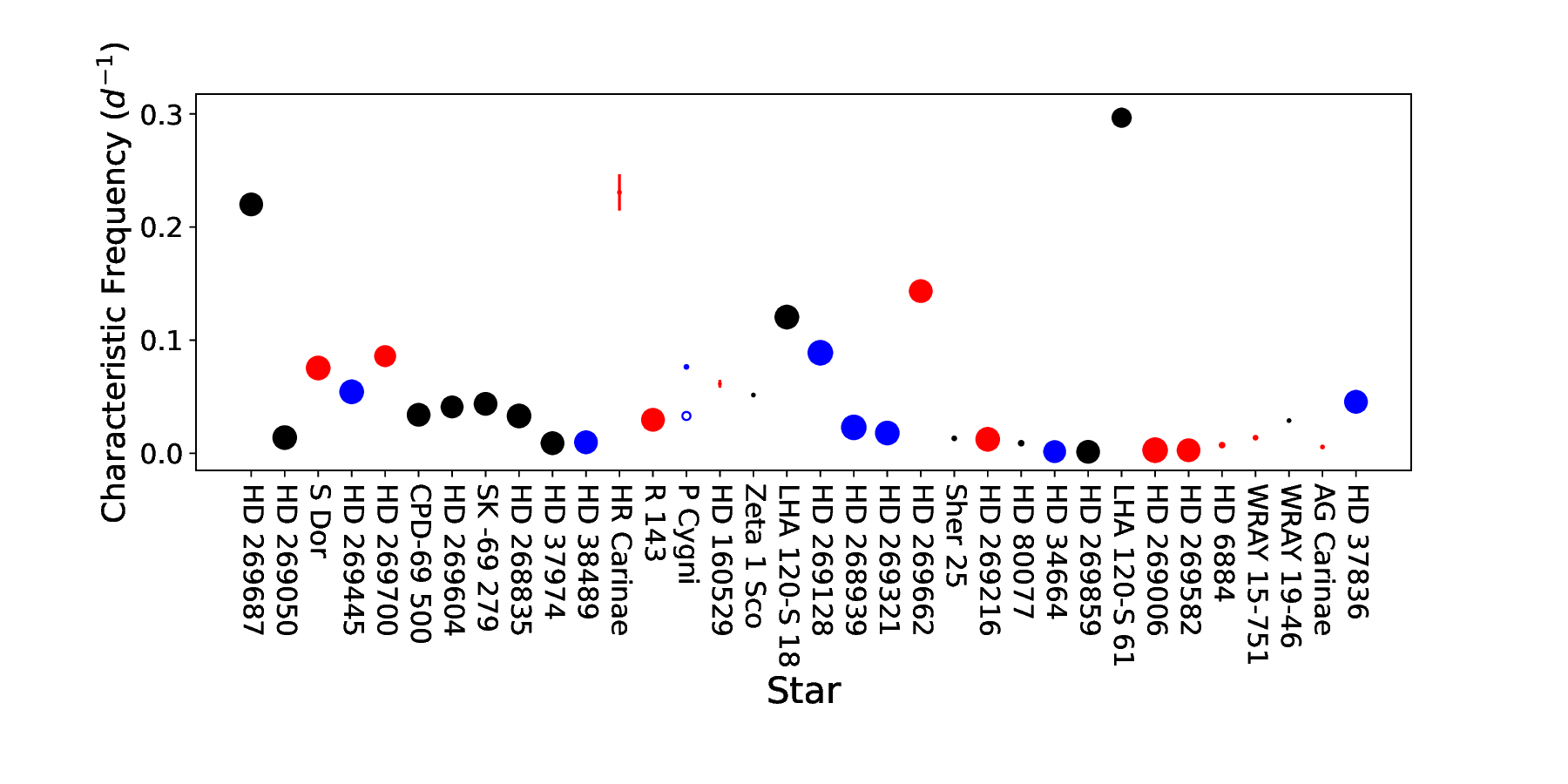}
\includegraphics[width=0.8\columnwidth]{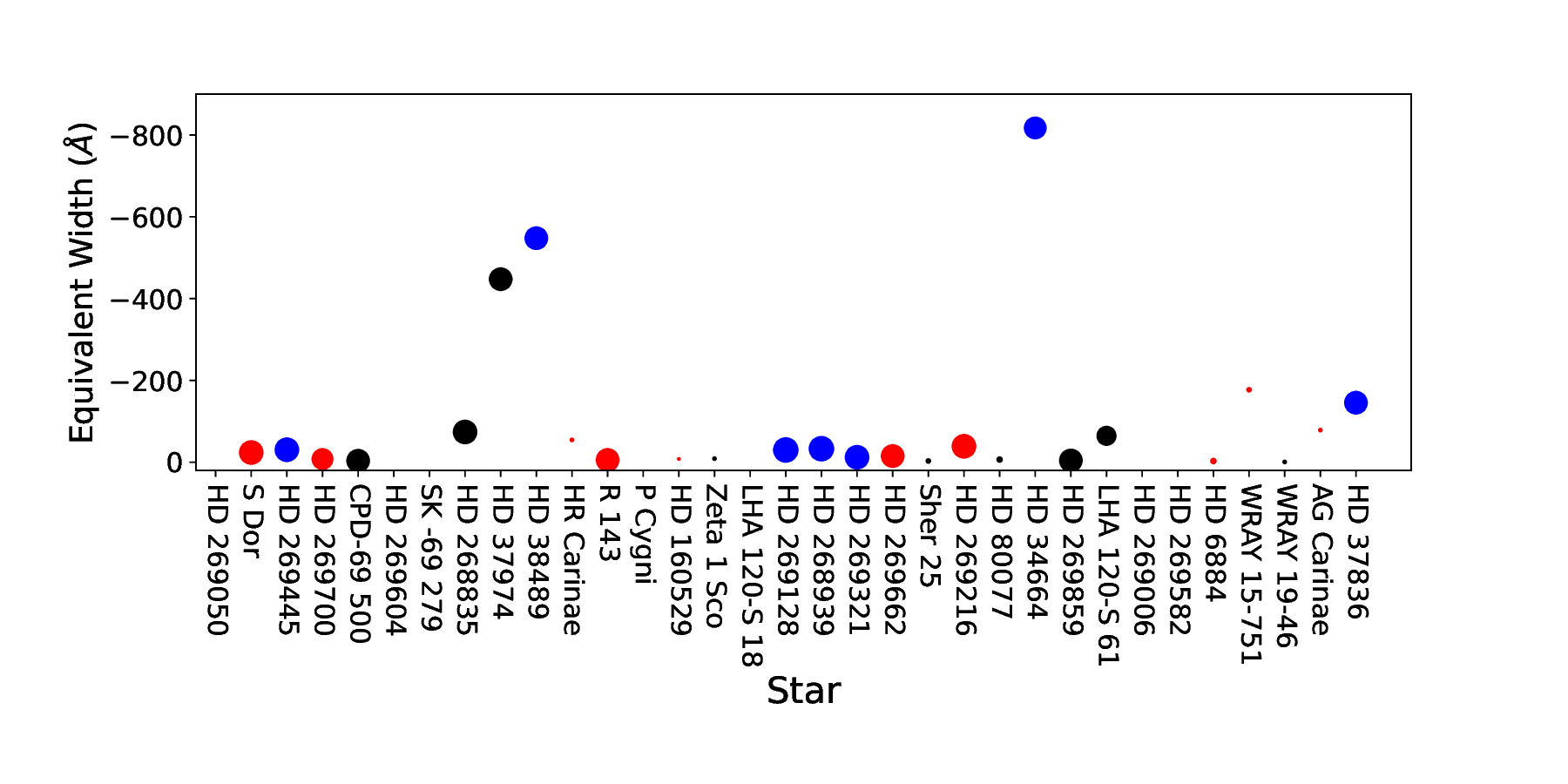}
\caption{The red noise amplitude (top), characteristic frequency (middle), and equivalent widths (bottom) for each LBV. In this display, we plot the points in order of increasing red noise amplitude and have made the other two plots match in order. The type of LBV is shown by dot color --- red for strong-active, blue for weak-active, and black for candidate/dormant. The open dot shows the values that \citet{2022MNRAS.509.4246E} measured with their complementary analysis of P Cygni.
\label{fig:ampsort}}
\end{figure}

% \begin{figure}[ht!]
% %\centering
% \caption{This graph shows the amplitude (left), characteristic frequency (center), and logarithmic equivalent widths for each LBV. The type of LBV is shown by dot color --- red for strong-active, blue for weak-active, and black for candidate/dormant. The open dot shows the values that \citet{2022MNRAS.509.4246E} got for their values of P Cygni.
% \label{fig:amp-freq-ha}}
% \end{figure}

% \begin{figure}[ht!]
% %\centering
% \caption{This graph shows the amplitude (left), characteristic frequency (center), and logarithmic equivalent widths for each LBV. The type of LBV is shown by dot color --- red for strong-active, blue for weak-active, and black for candidate/dormant. The open dot shows the values that \citet{2022MNRAS.509.4246E} got for their values of P Cygni.
% \label{fig:amp-freq-ha}}
% \end{figure}

\begin{figure}[ht!]
%\centering
\includegraphics[width=0.8\columnwidth]{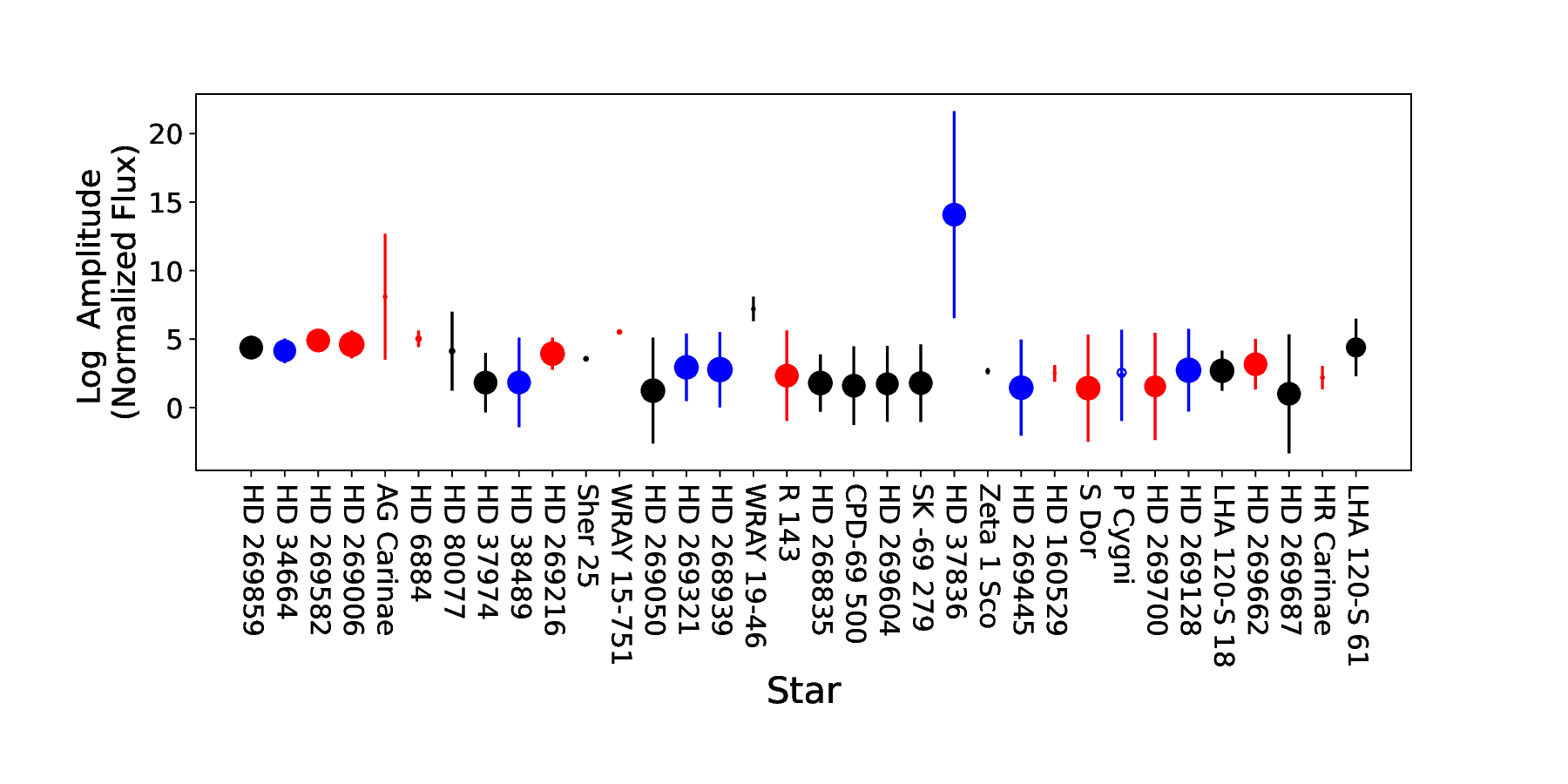}
\includegraphics[width=0.8\columnwidth]{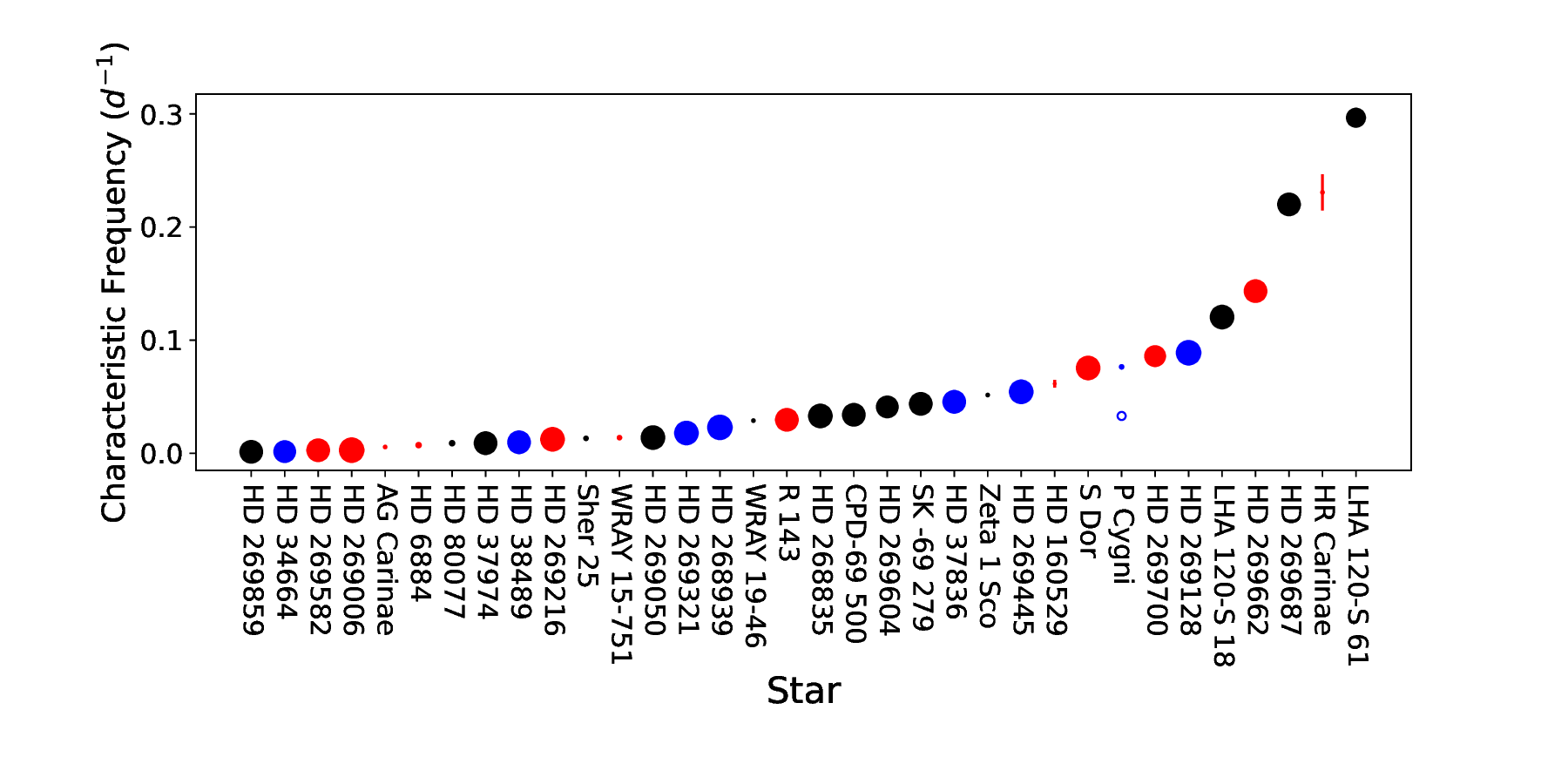}
\includegraphics[width=0.8\columnwidth]{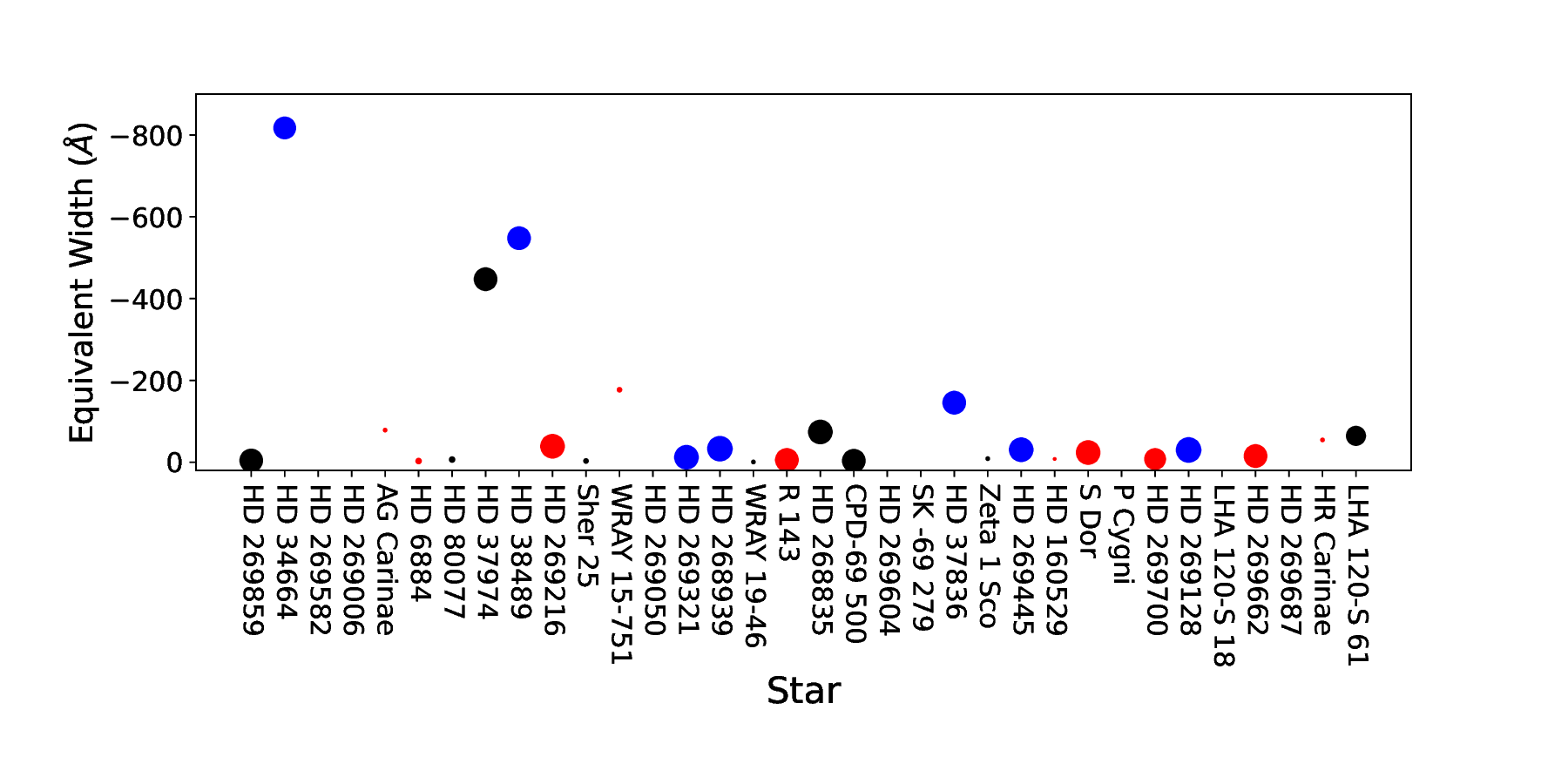}
\caption{The red noise amplitude (top), characteristic frequency (middle), and equivalent widths (bottom) for each LBV. In this display, we plot these in order of increasing characteristic frequency. The type of LBV is shown by dot color --- red for strong-active, blue for weak-active, and black for candidate/dormant. The open dot shows the values that \citet{2022MNRAS.509.4246E} got for their values of P Cygni.
\label{fig:freqsort}}
\end{figure}

% \begin{figure}[ht!]
% %\centering
% \caption{This graph shows the amplitude (left), characteristic frequency (center), and logarithmic equivalent widths for each LBV. The type of LBV is shown by dot color --- red for strong-active, blue for weak-active, and black for candidate/dormant. The open dot shows the values that \citet{2022MNRAS.509.4246E} got for their values of P Cygni.
% \label{fig:amp-freq-ha}}
% \end{figure}

% \begin{figure}[ht!]
% %\centering
% \caption{This graph shows the amplitude (left), characteristic frequency (center), and logarithmic equivalent widths for each LBV. The type of LBV is shown by dot color --- red for strong-active, blue for weak-active, and black for candidate/dormant. The open dot shows the values that \citet{2022MNRAS.509.4246E} got for their values of P Cygni.
% \label{fig:amp-freq-ha}}
% \end{figure}

\begin{figure}[ht!]
%\centering
\includegraphics[width=0.8\columnwidth]{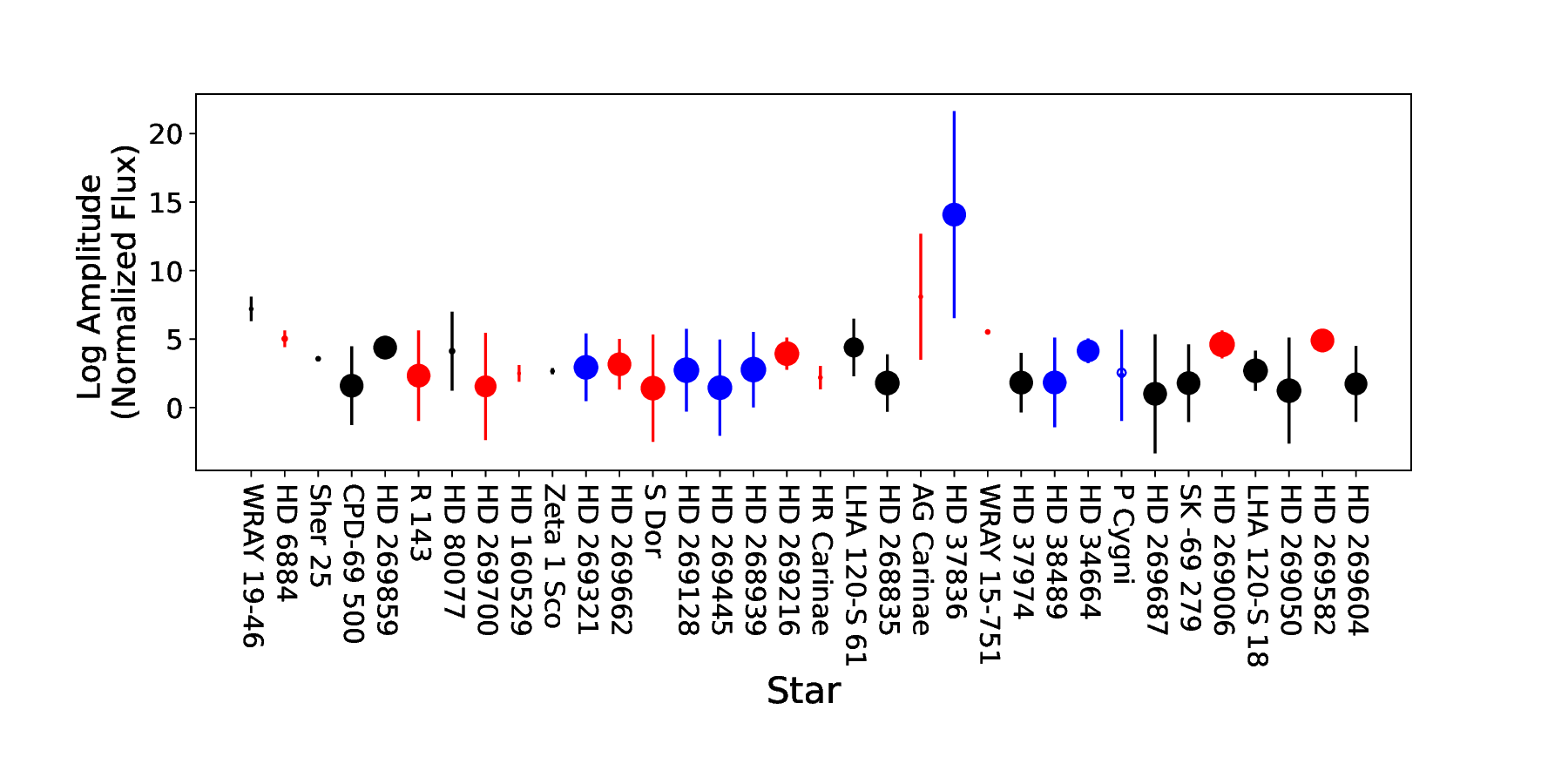}
\includegraphics[width=0.8\columnwidth]{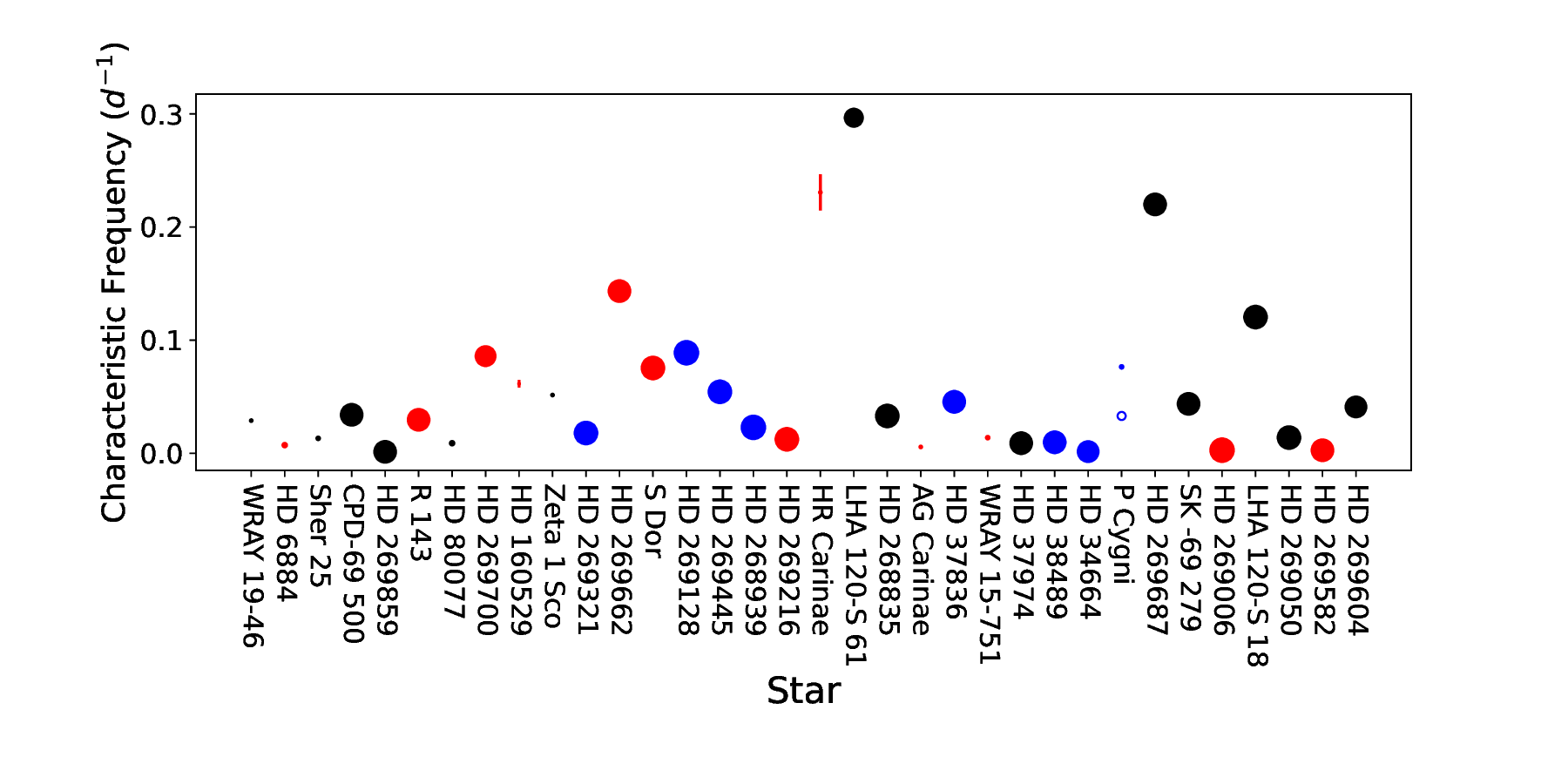}
\includegraphics[width=0.8\columnwidth]{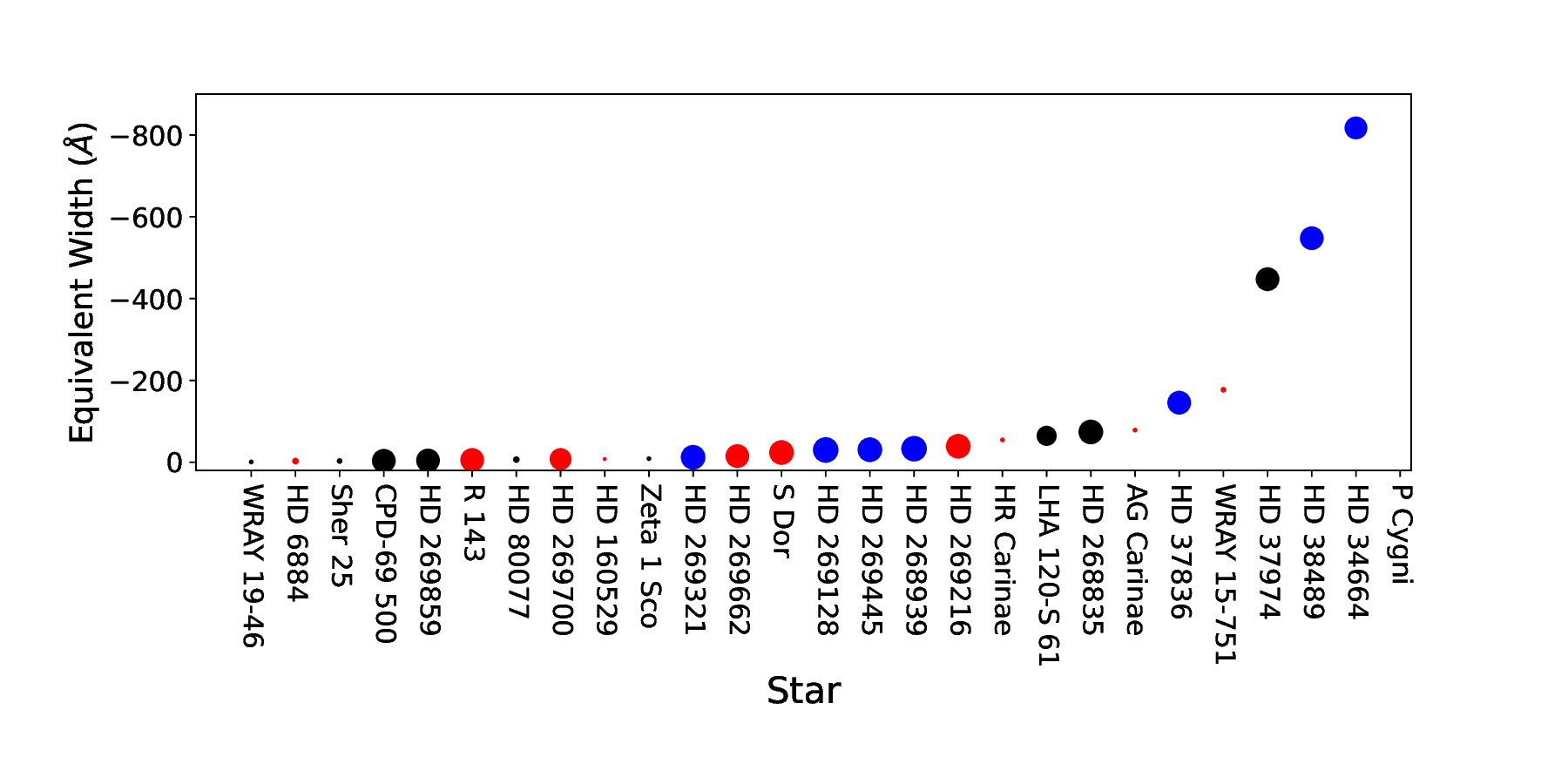}
\caption{The red noise amplitude (top), characteristic frequency (middle), and equivalent widths (bottom) for each LBV. In this display, we plot these in order of increasing H$\alpha$ strength. The type of LBV is shown by dot color --- red for strong-active, blue for weak-active, and black for candidate/dormant. The open dot shows the values that \citet{2022MNRAS.509.4246E} got for their values of P Cygni.
\label{fig:hasort}}
\end{figure}

% \begin{figure}[ht!]
% %\centering

% \caption{This graph shows the amplitude (left), characteristic frequency (center), and logarithmic equivalent widths for each LBV. The type of LBV is shown by dot color --- red for strong-active, blue for weak-active, and black for candidate/dormant. The open dot shows the values that \citet{2022MNRAS.509.4246E} got for their values of P Cygni.
% \label{fig:amp-freq-ha}}
% \end{figure}

% \begin{figure}[ht!]
% %\centering
% \caption{This graph shows the amplitude (left), characteristic frequency (center), and logarithmic equivalent widths for each LBV. The type of LBV is shown by dot color --- red for strong-active, blue for weak-active, and black for candidate/dormant. The open dot shows the values that \citet{2022MNRAS.509.4246E} got for their values of P Cygni.
% \label{fig:amp-freq-ha}}
% \end{figure}

\begin{figure}[ht!]
\centering
\includegraphics[width=\columnwidth]{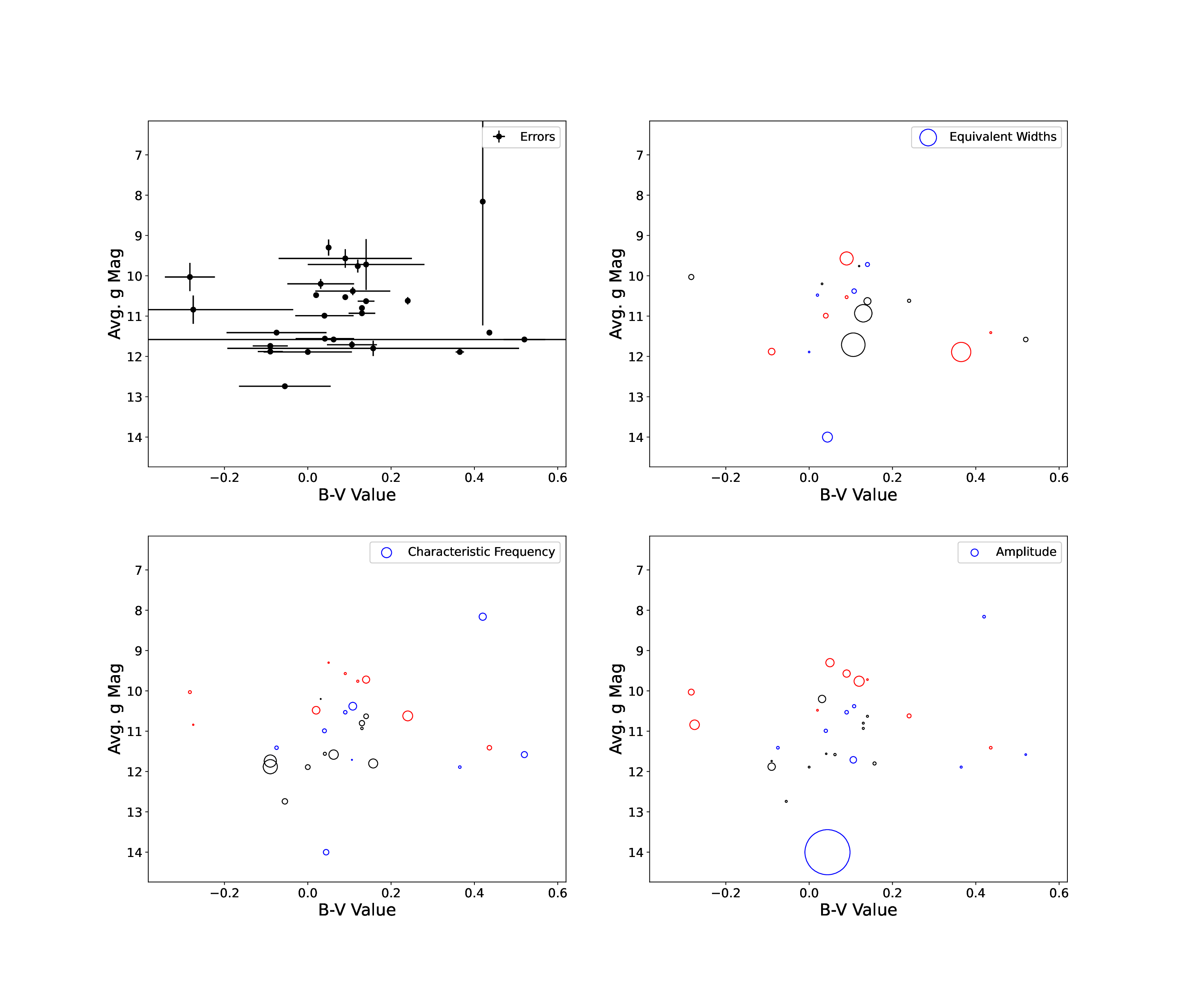}
\caption{This graph shows the B-V color for each LMC target compared to their average $g$ magnitudes. {Errors come from the standard deviation of the $g-$magnitudes from ASAS-SN survey and from the combination of errors in the reported $B$ and $V$ magnitudes.} From top to bottom, the size of the markers are proportional to each stars' equivalent width, characteristic frequency, and red noise amplitude.
\label{fig:HR}}
\end{figure}

\section{Discussion}

With the Fourier parameters from our large sample of Galactic and LMC/SMC LBVs, we are able to search for correlations between parameters to explore if there are indications to what drives the variability of these enigmatic stars. We begin this exploration with Figs.~\ref{fig:ampsort},~\ref{fig:freqsort}, and ~\ref{fig:hasort}. In Fig.~\ref{fig:ampsort}, we plot the red noise amplitude ($\alpha_0$) sorted by increasing strength. We also made the size of the symbol proportional to the number of \textit{TESS} sectors used to minimize any bias we may have from Fourier transforms being calculated with varying numbers of measurements and temporal coverage. We then plotted the characteristic frequency and equivalent widths of H$\alpha$ in separate panels but with the same star order in Fig.~\ref{fig:ampsort}. 

We repeated this ordering of the fit parameters for both the characteristic frequency and the equivalent width of H$\alpha$, shown in Figs.~\ref{fig:freqsort}, and ~\ref{fig:hasort} respectively. We find that there is no discernible pattern in the other measurements compared to the ordered parameter. In addition to ordering these points, we identified each source as a strong-active, weak-active, dormant, or candidate LBV using the criteria and classification from \citet{2001A&A...366..508V} when possible. Often, a stronger H$\alpha$ strength is somewhat correlated with hotter LBVs, as seen with S Doradus \citep{2010ATel.2560....1R}, so a lack of correlation here is also a proxy for the photospheric temperature.

Beyond these searches for correlations, we also wanted to explore if there were trends amongst the stars as situated on a color-magnitude diagram. In Fig.~\ref{fig:HR}, we show the stars from the LMC on an color-magnitude diagram with the average ASAS-SN $g$-magnitude as a proxy for the luminosity of these stars and the $B-V$ color index from the SIMBAD database for a proxy of temperature. This may not be the best way to place these stars on a diagram representing the H-R diagram, but it should be close given these stars are all in the LMC so should experience similar extinction and lie at the same distance from us. In each of the three plots, we show the measured value of the equivalent width of H$\alpha$, red noise amplitude and frequency in a symbol that has a size proportional to the measured value. We find no correlation in the placement of the stars with the measured quantities.

With no correlation found between the distinct parameters or the stars' placement in a color-magnitude diagram either through curve fitting or through visual inspection, we can ask two important questions. First, what does this imply for the driving mechanisms for the variability of LBVs? Second, how do the parameters vary based on the different types of LBVs, especially the strong-active, weak-active, and dormant/candidate types?

\citet{2022MNRAS.509.4246E} examined a prototype of LBVs, the weak-active P Cygni, with five years of precision time-series photometry. Our results on the same dataset provided similar measurements as those by \citet{2022MNRAS.509.4246E} with our comparable methodology. The driving mechanisms for P Cygni could be a type of sub-surface convection driven by helium opacity, as modeled by \citet{2018Natur.561..498J}. Alternatively, \citet{2022MNRAS.509.4246E} interpreted the stochastic variations of P Cygni could be the result of internal gravity waves which show up with lower characteristic frequencies and higher red noise amplitudes than those of lower-luminosity B supergiants. Both of these interpretations could explain our results. {\citet{2020A&A...640A..36B} examined a variability of main-sequence OB stars and showed that internal gravity waves impact the Fourier properties. In this analysis, the characteristic frequency, $\nu_{\rm char}$, decreases with increasing luminosity and decreasing effective temperature, while the amplitude, $\alpha_0$, increases with increasing
luminosity and decreasing temperature. \citet{2022MNRAS.509.4246E} used this in their analysis of the \textit{BRITE} light curve of P Cygni speculating that the variability they observed was due to internal gravity waves. However, \citet{2022MNRAS.509.4246E} also show that the variations of this star could be driven by sub-surface convection as detailed in the models of \citet{2018Natur.561..498J}. }

Another surprising finding from our analysis was the lack of a clear boundary between dormant/candidate, weak-active and strong-active LBVs. This is clearly seen in Figs.~\ref{fig:ampsort}, ~\ref{fig:freqsort}, and ~\ref{fig:hasort} where there is no clear trend of the points changing color representing the different types of LBVs. This could be because of a few reasons. While the strong-active LBVs show strong long-term behavior (e.g., Fig.~\ref{ASASSN}), the \textit{TESS} light curves still are predominantly sampling the short-period ($\lesssim 1$ month) variations, especially for the Galactic LBVs which suffer from much less time coverage with \textit{TESS}. Thus, the main finding of this work is the confirmation of the absence of specific trends based on luminosity traced with the $g-$magnitude, temperature traced by $B-V$, or dependent on other properties of the stochastic variability. Therefore it seems that the $\alpha$ Cygni type variations of these stars are an extension of those seen amongst the hot supergiants. 

With the $\alpha$ Cygni variations seen to be similar to, or an extension of, the hot supergiants, we can speculate that the long- and short- S Doradus cycles are extensions of these same behaviors. Thus, stars that reside in the appropriate part of the H-R diagram may all be LBVs without the observation of eruptions or the long- or short- S Doradus cycles. This would greatly simplify the classification of LBVs as the drivers {of the short-term variability that we probe here} seem independent of the temperatures, luminosities, or the red noise amplitudes and time-scales of variability. One way to fully explore this will be to do a similar study to this one once \textit{TESS} has surveyed the sky for at least a decade, or perhaps even with the less-precise ground-based data from the ASAS-SN survey in the future. 

\section{Conclusions}

We presented an analysis of the Fourier properties of the photometric variations of Galactic and LMC/SMC LBVs observed with \textit{TESS}. The \textit{TESS} photometry and Fourier properties allow us to measure the short-term $\alpha$ Cygni-type variations, which could help us model the driving mechanisms of these stars similar to how \textit{TESS} and \textit{K2} observations informed the asteroseismology of B-type supergiants \citep{2019NatAs...3..760B}. Suggested models of these variations, as seen in the P Cygni, are likely either internal gravity waves 
\citep[e.g.,][]{2020A&A...640A..36B} or sub-surface convection as modeled by \citet{2018Natur.561..498J}. As the timescales sampled by the microvariations are similar to those in the literature, we did not explore the strange mode instability as \citet{1998A&A...335..605L} discussed the time scale observed tends to be too long for strange modes. None of the Fourier properties correlated with intrinsic color or the H$\alpha$ strengths of these stars, showing that the $\alpha$ Cygni variations for these stars exist in some basic parameter space. These lack of correlations confirm earlier findings of \citet{2021MNRAS.502.5038N} who explored these parameters for a smaller set of LBVs, {whereas our sample includes the LMC and SMC LBVs greatly expanding upon this work}.
As such, we speculate that the LBV classification may not need a variability criterion but rather is a phase of an extreme supergiant in the H-R diagram. 
Future studies could investigate the Fourier properties of these stars with ground-based photometric time-series. Comparing these properties to the Fourier properties of spectroscopic measurements could help elucidate how the stars are varying and such could help distinguish between the sub-surface convection and internal gravity waves. 

\section*{acknowledgments}

We thank the anonymous referee for constructive feedback. We also wish to thank Dominic Bowman, Yan-Fei Jiang, Phil Massey, and Andrea Mehner for comments that helped to improve this manuscript.
B.S. and M.B. are thankful for funding from the Embry-Riddle Aeronautical University Undergraduate Research Institute through their IGNITE program. B.S. also was funded through the NASA Space Grant program. N.D.R. is grateful for funding through NASA Awards 80NSSC23K1049 and 80NSSC24K0229.
P.B. and E.A. participated in this research through the BASIS Prescott Senior capstone program, and we are thankful to their guidance counselors for helping connect them to our group.

Our spectroscopy was obtained through NOIRLab with program 2022A-177963. This research has
used data from the CTIO/SMARTS 1.5m telescope, which is operated as part of the SMARTS
Consortium by RECONS (www.recons.org) members T. Henry, H. James, W.-C. Jao and
L. Paredes. At the telescope, observations were carried out by R. Aviles and R. Hinojosa. This paper includes data collected by the TESS mission. Funding for the TESS mission is provided by the NASA's Science Mission Directorate.

%\end{acknowledgments}


\vspace{5mm}
\facilities{TESS, 
CTIO:1.5m}


\software{astropy \citep{2013A&A...558A..33A,2018AJ....156..123A},
Period04 \citep{2005CoAst.146...53L}, 
eleanor \citep{2019PASP..131i4502F}}

\bibliography{sample631}{}
\bibliographystyle{aasjournal}



\end{document}